\documentclass[twocolumn]{aastex631}

\usepackage{comment}
\usepackage{xspace}

\newcommand{\um}{\text{\textmu m}\xspace}
\newcommand{\Mjup}{\rm{M}_{\rm{Jup}}\xspace}
\newcommand{\Rjup}{\rm{R}_{\rm{Jup}}\xspace}
\newcommand{\bpic}{$\beta$~Pic\xspace}
\newcommand{\bpicb}{$\beta$~Pic~b\xspace}
\newcommand{\bpicc}{$\beta$~Pic~c\xspace}
\newcommand{\apic}{$\alpha$~Pic\xspace}

\newcommand{\rev}[1]{{#1}}

\begin{document}

\title{JWST-TST High Contrast: JWST/NIRCam observations of the young giant planet \bpicb}

\author[0000-0003-2769-0438]{Jens Kammerer}
\affiliation{European Southern Observatory, Karl-Schwarzschild-Straße 2, 85748 Garching, Germany}
\affiliation{Space Telescope Science Institute, 3700 San Martin Drive, Baltimore, MD 21218, USA}

\author[0000-0002-6964-8732]{Kellen Lawson}
\affiliation{NASA Goddard Space Flight Center, Greenbelt, MD 20771, USA}

\author[0000-0002-3191-8151]{Marshall D. Perrin}
\affiliation{Space Telescope Science Institute, 3700 San Martin Drive, Baltimore, MD 21218, USA}

\author[0000-0002-4388-6417]{Isabel Rebollido}
\affiliation{European Space Agency (ESA), European Space Astronomy Centre (ESAC), Camino Bajo del Castillo s/n, 28692 Villanueva de la Ca\~nada, Madrid, Spain}
\affiliation{Centro de Astrobiolog\'ia (CAB CSIC-INTA) ESAC Campus Camino Bajo del Castillo, Villanueva de la Cañada, 28692 Madrid, Spain}

\author{Christopher C. Stark}
\affiliation{NASA Goddard Space Flight Center, Greenbelt, MD 20771, USA}

\author[0000-0002-5823-3072]{Tomas Stolker}
\affiliation{Leiden Observatory, Leiden University, Niels Bohrweg 2, 2333 CA Leiden, The Netherlands}

\author[0000-0001-8627-0404]{Julien H. Girard}
\affiliation{Space Telescope Science Institute, 3700 San Martin Drive, Baltimore, MD 21218, USA}

\author{Laurent Pueyo}
\affiliation{Space Telescope Science Institute, 3700 San Martin Drive, Baltimore, MD 21218, USA}

\author[0000-0001-6396-8439]{William O. Balmer}
\affiliation{Department of Physics \& Astronomy, Johns Hopkins University, 3400 N. Charles Street, Baltimore, MD 21218, USA}
\affiliation{Space Telescope Science Institute, 3700 San Martin Drive, Baltimore, MD 21218, USA}

\author[0000-0002-5885-5779]{Kadin Worthen}
\affiliation{Department of Physics \& Astronomy, Johns Hopkins University, 3400 N. Charles Street, Baltimore, MD 21218, USA}

\author[0000-0002-8382-0447]{Christine Chen}
\affiliation{Space Telescope Science Institute, 3700 San Martin Drive, Baltimore, MD 21218, USA}

\author[0000-0001-7827-7825]{Roeland P. van der Marel}
\affiliation{Space Telescope Science Institute, 3700 San Martin Drive, Baltimore, MD 21218, USA}
\affiliation{Department of Physics \& Astronomy, Johns Hopkins University, 3400 N. Charles Street, Baltimore, MD 21218, USA}

\author[0000-0002-8507-1304]{Nikole K. Lewis}
\affiliation{Department of Astronomy and Carl Sagan Institute, Cornell University, 122 Sciences Drive, Ithaca, NY 14853, USA}

\author[0000-0002-4479-8291]{Kimberly Ward-Duong}
\affiliation{Department of Astronomy, Smith College, Northampton MA 01063, USA}

\author[0000-0003-3305-6281]{Jeff A. Valenti}
\affiliation{Space Telescope Science Institute, 3700 San Martin Drive, Baltimore, MD 21218, USA}

\author[0000-0003-4003-8348]{Mark Clampin}
\affiliation{Astrophysics Division, Science Mission Directorate, NASA Headquarters, 300 E Street SW, Washington, DC 20546, USA}

\author{C. Matt Mountain}
\affiliation{Association of Universities for Research in Astronomy, 1331 Pennsylvania Avenue NW Suite 1475, Washington, DC 20004, USA}

\begin{abstract}
We present the first \emph{JWST}/NIRCam observations of the directly-imaged gas giant exoplanet \bpicb. Observations in six filters using NIRCam’s round coronagraphic masks provide a high signal-to-noise detection of \bpicb and the archetypal debris disk around \bpic over a wavelength range of $\sim1.7$--$5~\um$. This paper focuses on the detection of \bpicb and other potential point sources in the NIRCam data, following a paper by Rebollido et al.\ which presented the NIRCam and MIRI view of the debris disk around \bpic.
We develop and validate approaches for obtaining accurate photometry of planets in the presence of bright, complex circumstellar backgrounds. By simultaneously fitting the planet's PSF and a geometric model for the disk, we obtain planet photometry that is in good agreement with previous measurements from the ground. The NIRCam data supports the cloudy nature of \bpicb's atmosphere and the discrepancy between its mass as inferred from evolutionary models and the dynamical mass reported in the literature. We further identify five additional localized sources in the data, but all of them are found to be background stars or galaxies based on their color or spatial extent. We can rule out additional planets in the disk midplane above $1~\Mjup$ outward of 2~arcsec ($\sim40$~au) and away from the disk midplane above $0.05~\Mjup$ outward of 4~arcsec ($\sim80$~au). The inner giant planet \bpicc remains undetected behind the coronagraphic masks of NIRCam in our observations.
\end{abstract}


\section{Introduction}
\label{sec:introduction}

The archetypal member of the nearby \bpic moving group \citep{zuckerman2001}, the A-type star \bpic itself, was the first star around which a circumstellar disk was directly imaged \citep{smith1984}, and one of the first to have a planet directly imaged orbiting it \citep{lagrange2009}. Already in the discovery paper by \citet{smith1984}, the disk was found to be extended to more than 400~au from the star and to be viewed from the Earth in an edge-on configuration. The shape of the disk was suspected to be influenced by planet formation, a hypothesis that was later supported by a variety of observational and theoretical studies \citep{burrows1995,kalas1995,mouillet1997,heap2000,augereau2001,wahhaj2003,golimowski2006}. In 2009, a gas giant exoplanet was imaged in the inner gap of the debris disk. Discovered by \citet{lagrange2009} using VLT/NACO observations in the L'-band ($\sim3.8~\um$), \bpicb was found to orbit its young \citep[$\sim18.5$~Myr,][]{miret-roig2020} host star in a highly inclined $\sim10$~au orbit, coplanar with the debris disk. At this orbital separation, which is similar to that of Saturn, \bpicb remains one of the closest-in directly imaged exoplanets to date \citep[e.g.,][]{currie2022} and represents a prime testbed for studying the formation and early evolution of gas giant planets on Solar System-like scales.

Thanks to its proximity \citep[distance $\sim$ 19.44~pc,][]{vanleeuwen2007} and apparent brightness, \bpicb quickly became one of the most extensively studied exoplanets. Observations from $\sim1$--$5~\um$ probing the thermal emission of the planet are discussed in numerous publications \citep{lagrange2009,quanz2010,bonnefoy2011,currie2011,bonnefoy2013,currie2013,absil2013,males2014,baudino2015,morzinski2015,stolker2020}, and the first low-resolution J-band spectroscopy (obtained with Gemini/GPI) was presented in \citet{bonnefoy2014}, later complemented by GPI H- and K-band spectroscopy \citep{morzinski2015,chilcote2017}. Based on the aforementioned studies, \bpicb was found to have an effective temperature of $\sim1700$~K, a radius of $\sim1.4$--$1.5~\Rjup$, and a surface gravity of $\log g \sim 4.2$~dex estimated from the bolometric luminosity of the planet and the age of the system using warm- or hot-start evolutionary models with moderate to high formation entropies $\gtrsim9.75~k_{\rm{B}}/\rm{baryon}$ \citep[e.g.,][]{spiegel2012}. Its spectral type lies somewhere between L0--T0. The red L-M color of \bpicb further suggests a cloudy atmosphere \citep{currie2013,bonnefoy2013,stolker2020} with possibly even thicker clouds or stronger water absorption than predicted by common cloudy atmosphere models such as AMES-Dusty \citep{baraffe2003} or DRIFT-PHOENIX \citep{helling2008}.

\rev{The age of \bpic and the \bpic moving group has been the subject of numerous publications. \citet{couture2023} compile an age range of $\sim11$--26~Myr from the literature, with their own dynamical traceback age estimate of $20.4\pm2.5$~Myr. A more recent study from \citet{lee2024} also finds a kinematic traceback age ($16.3^{+3.4}_{-2.1}$~Myr) that is younger than the age determined from lithium depletion boundary ($\sim24$~Myr). As noted in \citet{couture2023}, the dynamical traceback age is the time since when the members of a group became gravitationally unbound, which might be systematically different from the time of their formation if young stars remain bound by interstellar gas and dust for a few million years after their formation. However, studying potential systematics in different stellar age dating methods shall not be the subject of this work. Here, we assume the dynamical traceback age of $18.5$~Myr inferred by \citet{miret-roig2020}, which lies somewhere in between the estimates from \citet{lee2024}, \citet{crundall2019}, and \citet{couture2023}, but we note that this age is slightly younger than the overall reported age average. We recognize that assuming a slightly younger age might yield a slightly smaller planet mass and a slightly higher expected planet luminosity based on evolutionary models which will be discussed in Section~\ref{sec:discussion}.}

Using $R \sim 5000$ VLT/SINFONI integral field spectroscopy and cross-correlation techniques, \citet{hoeijmakers2018} directly detected water and CO in the atmosphere of \bpicb, albeit without the possibility to derive abundance measurements. Two years later and taking advantage of the spatial resolution of long-baseline interferometry to separate the light of \bpicb from contamination by its host star and the debris disk, \citet{gravity2020} obtained high signal-to-noise ratio (SNR) $R \sim 500$ K-band spectroscopy of the young gas giant planet, clearly showing CO absorption features at $\sim2.3~\um$. With the help of atmospheric forward models and spectral retrievals, they were able to derive a subsolar C/O ratio of $\sim0.43$ for the atmosphere of \bpicb, suggesting strong volatile enrichment during its formation process if the host star \bpic is assumed to be of similar composition as the Sun. We note that measuring the C/O abundance ratio for \bpic is difficult, but it can be done for other stars in the \bpic moving group which can serve as a proxy for the chemical composition of \bpic \citep[e.g.,][]{reggiani2024}. While the C/O abundance ratio of the host star \bpic can hence be expected to be $\sim$solar or supersolar, the subsolar C/O ratio of \bpicb has recently been challenged by \citet{kiefer2024} who found that with a more careful treatment of telluric and stellar features, the SINFONI spectrum of \bpicb also supports a $\sim$solar C/O ratio for the planet. \rev{\citet{landman2024}, however, also detect water and CO in \bpicb's atmosphere and confirm its subsolar C/O ratio using high-resolution ($R \sim 100'000$) spectroscopy with the upgraded VLT/CRIRES+ instrument, updating the result that was earlier obtained by \citet{snellen2014} with the original CRIRES instrument.}

More recently, a second gas giant planet was detected in the \bpic system based on extensive radial velocity monitoring of the host star with the HARPS spectrograph at ESO's 3.6~m Telescope \citep{lagrange2019}. Using interferometric observations with the VLTI/GRAVITY instrument, \citet{nowak2020} were able to directly confirm this inner planet \bpicc at an orbital separation of only $\sim3$~au. 

Combined together, the precise astrometry of the two planets measured with GRAVITY, the long-term radial velocity monitoring with HARPS, and the Hipparcos-Gaia proper motion anomaly of the system \citep{brandt2021,kervella2022} yield dynamical mass constraints of $\sim9\pm2~\Mjup$ and $\sim8\pm1~\Mjup$ for \bpicb and \bpicc, respectively \citep{grandjean2019,nielsen2020,nowak2020,brandtetal2021}. \citet{lacour2021} were even able to measure the dynamical mass of \bpicc from the astrometric deflection that it causes on the orbit of \bpicb, similar to the discovery of Neptune from the orbital motion of Uranus by Le Verrier in 1846. The availability of model-independent dynamical masses makes the \bpic planets benchmark objects for probing gas giant planet formation and evolution.

\subsection{Programmatic context}
\label{sec:programmatic_context}

This paper is part of a series by the \emph{JWST} Telescope Scientist Team (JWST-TST)\footnote{\url{https://www.stsci.edu/~marel/jwsttelsciteam.html}}, which uses Guaranteed Time Observer (GTO) time awarded by NASA in 2003 (PI: M.~Mountain) for studies in three different subject areas: (a) Transiting Exoplanet Spectroscopy (lead: N.~Lewis); (b) Exoplanet and Debris Disk High-Contrast Imaging (lead: M.~Perrin); and (c) Local Group Proper Motion Science (lead: R.~van der Marel). Here, we present new results in the second area; previously reported results across these areas include \citet{grant2023} and Gressier et al. (in prep.); \citet{ruffio2023} and \citet{rebollido2024}; and \citet{libralato2023}, respectively. A common theme of these investigations is the desire to pursue and demonstrate science for the astronomical community at the limits of what is made possible by the exquisite optics and stability of \emph{JWST}.

In this paper, we continue to present the first coronagraphic observations of the \bpic system with the \emph{James Webb Space Telescope}'s \citep[\emph{JWST},][]{gardner2006,gardner2023} Near-InfraRed Camera \citep[NIRCam,][]{rieke2003,rieke2023}, spanning a wavelength range from $\sim1.7$--$5~\um$. We previously reported in \citet{rebollido2024} on the edge-on debris disk around \bpic as seen in these same NIRCam data, along with coronagraphic observations at 15.5 and $23~\um$ obtained with the Mid-InfraRed Instrument \citep[MIRI,][]{rieke2015,wright2015,wright2023}, which led to the discovery of a dramatic curved ``tail'' of dust that appears to extend sharply away from the plane of the disk, and may be unbound debris from a recent major collision of large planetesimals. 

In this work, we turn our attention to the planets within the \bpic system. The outer giant planet \bpicb is detected in all six observed NIRCam filters close to its greatest orbital elongation. The inner giant planet \bpicc remains undetected behind the coronagraphic masks of NIRCam. Both known planets \bpicb and \bpicc remain undetected in the MIRI coronagraphy data due to the tremendously bright debris disk in the mid-infrared, so that we choose to not further discuss those MIRI data here, except for one brief usage in Section~\ref{sec:limits_for_other_companions_in_the_bpic_system} for evaluating the nature of additional candidate companions detected in the NIRCam data. However, there are also MIRI Medium Resolution Spectroscopy \citep[MRS,][]{wells2015,argyriou2023} observations of the \bpic system (program 1294, PI: C.~Chen) which detect the outer planet \bpicb at longer wavelengths ($\sim5$--$7~\um$, \citealp{worthen2024}); we make use of their spectrum in joint analyses below.

The remainder of this paper is organized as follows. Section~\ref{sec:observations_and_data_reduction} presents the observations and data reduction, including two independent approaches to PSF subtraction. Section~\ref{sec:results} presents the measured photometry and astrometry of \bpicb, and detection limits on lower-mass planets in the outer system. In Section~\ref{sec:discussion}, we discuss the implications for \bpicb's atmospheric and bulk properties and its evolutionary status using the NIRCam measurements, prior literature values, and exoplanetary model grids. We summarize our findings and conclude in Section~\ref{sec:summary_and_conclusions}.

\section{Observations and data reduction}
\label{sec:observations_and_data_reduction}

\begin{table*}[!t]
    \centering
    \begin{tabular}{c c c c c c c}
        Target & Filter & Readout pattern & Dither pattern & $N_{\rm{ints}}$/$N_{\rm{groups}}$/$N_{\rm{frames}}$ & $T_{\rm{exp}}$ [s] & PA [deg] \\
        \hline
        \hline
        \bpic & F182M & RAPID & NONE & 90/4/1 & 1506.9 & 84.34 \\
        \bpic & F182M & RAPID & NONE & 90/4/1 & 1506.9 & 94.34 \\
        \apic & F182M & RAPID & 5-POINT-BOX & 10/4/1 & 837.2 & 76.46 \\
        \hline
        \bpic & F210M & RAPID & NONE & 90/4/1 & 1506.9 & 84.34 \\
        \bpic & F210M & RAPID & NONE & 90/4/1 & 1506.9 & 94.34 \\
        \apic & F210M & RAPID & 5-POINT-BOX & 10/4/1 & 837.2 & 76.46 \\
        \hline
        \bpic & F250M & BRIGHT2 & NONE & 80/10/2 & 1710.5 & 84.57 \\
        \bpic & F250M & BRIGHT2 & NONE & 80/10/2 & 1710.5 & 94.57 \\
        \apic & F250M & BRIGHT2 & 5-POINT-BOX & 8/10/2 & 855.2 & 76.39 \\
        \hline
        \bpic & F300M & BRIGHT2 & NONE & 80/10/2 & 1710.5 & 84.57 \\
        \bpic & F300M & BRIGHT2 & NONE & 80/10/2 & 1710.5 & 94.57 \\
        \apic & F300M & BRIGHT2 & 5-POINT-BOX & 8/10/2 & 855.2 & 76.39 \\
        \hline
        \bpic & F335M & SHALLOW4 & NONE & 35/10/4 & 1833.4 & 84.57 \\
        \bpic & F335M & SHALLOW4 & NONE & 35/10/4 & 1833.4 & 94.57 \\
        \apic & F335M & SHALLOW4 & 5-POINT-BOX & 4/10/4 & 1047.7 & 76.39 \\
        \hline
        \bpic & F444W & SHALLOW4 & NONE & 35/10/4 & 1833.4 & 84.57 \\
        \bpic & F444W & SHALLOW4 & NONE & 35/10/4 & 1833.4 & 94.57 \\
        \apic & F444W & SHALLOW4 & 5-POINT-BOX & 4/10/4 & 1047.7 & 76.39 \\
    \end{tabular}
    \caption{Summary of the \emph{JWST}/NIRCam observations of the young giant planet \bpicb presented in this paper (\emph{JWST} program ID 1411). The science target is \bpic and the PSF reference target is \apic. The observations were taken on 18 March 2023. The total exposure time $T_{\rm{exp}}$ sums up all five dither positions for the reference target.}
    \label{tab:observations}
\end{table*}

The \emph{JWST} data presented here \rev{were} taken as part of GTO program 1411\footnote{\url{https://www.stsci.edu/jwst/science-execution/program-information.html?id=1411}} (``Coronagraphy of the Debris Disk Archetype Beta Pictoris'', PI: C.~Stark) on 18 March 2023\footnote{These NIRCam observations were initially attempted on 22 December 2022, but failed due to a guiding issue. The observations were subsequently rescheduled roughly three months later, at a position angle rotated by approximately 90~degrees from the original plan.}. The program was mainly designed to characterize the disk around \bpic using NIRCam filters sensitive to the presence of water, CO ices, and organic tholins and MIRI filters probing the warm inner asteroid belt analogue and the cooler outer main disk. However, given the timing of \emph{JWST}'s launch, the well-known gas giant planet \bpicb happened to be close to its greatest orbital elongation during the observations, enabling us to confidently detect and characterize its atmosphere from $\sim1.7$--$5~\um$.

\subsection{Observational structure}
\label{sec:observational_structure}

The NIRCam observations targeted \bpic in six different filters from $\sim1.7$--$5~\um$, two of which were observed using NIRCam's short wavelength (SW) channel with a pixel scale of $\sim31$~mas (F182M, F210M) and four of which were observed using its long wavelength (LW) channel with a pixel scale of $\sim63$~mas (F250M, F300M, F335M, F444W). The F444W wide-band filter was included because it is the most sensitive to additional, yet-undiscovered planets in the system \citep[e.g.,][]{carter2023}. The SW observations were taken using NIRCam's 210R round coronagraphic mask; these were the first-ever science observations taken with SW channel coronagraphy.  The LW observations were taken using the 335R mask. These masks were chosen because they are the smallest available round coronagraphic masks providing the best performance at small angular separations while retaining access to the entire 360~deg field-of-view. We note that the observations were taken sequentially in the two channels, as parallel operation of the SW and LW channels was not yet enabled at the time.

The observations were conducted using the standard high-contrast imaging strategy of two rolls on the science target plus a PSF reference star. The science target \bpic was observed at two roll angles offset by a position angle (PA) of $\sim10$~deg from one another. This enables point-spread function (PSF) subtraction using angular differential imaging (ADI) techniques \rev{\citep{liu2004,marois2006}}, which are frequently used to detect faint companions  \citep[e.g.,][]{bowler2016,currie2022}. In addition, a PSF reference star (\apic) was observed in sequence with the science target using the 5-POINT-BOX\footnote{\url{https://jwst-docs.stsci.edu/jwst-near-infrared-camera/nircam-operations/nircam-dithers-and-mosaics/nircam-subpixel-dithers/nircam-small-grid-dithers}} small grid dither (SGD) pattern. This enables reference differential imaging (RDI), including classical PSF subtraction methods which are typically less prone to self- or oversubtraction of extended circumstellar structure such as the debris disk around \bpic \citep{lafreniere2007,soummer2012}. The SGD can be used to enhance the quality of the PSF subtraction (i.e., to suppress residual stellar speckles in the PSF-subtracted images) by sampling changes in the shape of the PSF as a function of the relative misalignment between the star and the coronagraphic mask \citep{soummer2014}. This is necessary because the target acquisition (TA) procedure for NIRCam coronagraphic imaging can have uncertainties of typically up to $\sim20$~mas as measured in flight \citep{girard2022,rigby2023}. Table~\ref{tab:observations} shows a summary of the NIRCam coronagraphic observations.

\subsection{Data reduction}
\label{sec:data_reduction}

All data were reduced with the \texttt{spaceKLIP}\footnote{\url{https://github.com/kammerje/spaceKLIP}} community pipeline \citep{kammerer2022,carter2023}. This pipeline was developed during \emph{JWST} commissioning and Early Release Science (ERS) and has since been updated and improved steadily. The data reduction process is briefly summarized in the following, with the individual steps described in more detail in the subsequent subsections.
\begin{enumerate}
    \item Reduce the raw (``uncal'') files with the \texttt{jwst}\footnote{\url{https://github.com/spacetelescope/jwst}} stage 1 and 2 pipelines \citep{bushouse2023} first to ``rateints'' and then to flux-calibrated ``calints'' files, with several custom adaptions specific to NIRCam coronagraphy data that are implemented in \texttt{spaceKLIP}.
    \item Use the \texttt{spaceKLIP} image processing library to clean bad pixels and align the PSFs in preparation for PSF subtractions.
    \item Perform PSF subtractions. Several complementary methods were employed, as described below. In particular, we used the \texttt{pyKLIP}\footnote{\url{https://bitbucket.org/pyKLIP/pyklip/src/master/}} community pipeline \citep{wang2015} invoked through \texttt{spaceKLIP} to perform PSF subtractions using projections on Karhunen-Lo\`eve eigenimages \citep[KLIP,][]{soummer2012}. We also applied Model-Constrained Reference Differential Imaging \citep[MCRDI,][]{lawson2022} as an alternative approach to help separate disk and planetary fluxes.
    \item For the KLIP subtractions, we use the \texttt{pyKLIP} forward-modeling routines \citep{pueyo2016} with NIRCam coronagraphy PSF models generated through \texttt{spaceKLIP} to extract companion astrometry and photometry and compute contrast curves.
\end{enumerate}

\subsubsection{Calibration of individual images}
\label{sec:calibration_of_individual_images}

In the first stage, we downloaded the ``uncal'' files from the MAST archive and processed them with the \texttt{Coron1Pipeline} and \texttt{Coron2Pipeline} within \texttt{spaceKLIP}. These two pipelines are customized implementations of the \texttt{jwst} \texttt{Detector1Pipeline} and \texttt{Image2Pipeline}. We used version 1.12.1 of the \texttt{jwst} pipeline and calibration context \texttt{jwst\_1174.pmap} for reducing the NIRCam data. As in \citet{carter2023}, we disabled the flagging of pixels that are diagonal neighbors to saturated pixels, skipped the pipeline's dark subtraction step, flagged a four pixels wide border around the edge of the subarrays to be used as ``pseudo'' reference pixels, and set the jump rejection threshold to 4. In addition, and as already discussed in \citet{rebollido2024}, we also employed a custom 1/f stripe noise correction step that is now implemented in \texttt{spaceKLIP} and yields a mostly cosmetic improvement. The output of this stage are flux-calibrated ``calints'' files in units of MJy/steradian. This reduction made use of the first in-flight photometric calibrations for NIRCam coronagraphy delivered in fall 2023\footnote{For example, reference file \texttt{jwst\_nircam\_photom\_0157.fits} available starting in CRDS context \texttt{jwst\_1146.pmap}.}. That on-sky flux calibration accounts for the reduced throughput of the NIRCam coronagraphic Lyot stops (pupil plane masks), but not for the attenuation of the focal plane masks for sources located at small angular separation from the observed target\footnote{\url{https://jwst-docs.stsci.edu/jwst-near-infrared-camera/nircam-instrumentation/nircam-coronagraphic-occulting-masks-and-lyot-stops}}. It also accounts for the wavelength-dependent transmission of the coronagraphic mask (COM) substrate \citep{krist2010}. \rev{The attenuation of the coronagraphic (focal plane) masks is accounted for later in the companion fitting step, when the model PSF of the companion is computed (see Section~\ref{sec:jwst_photometry}).}

In the second stage, we prepared the reduced images for processing with \texttt{pyKLIP}. First, we median-subtracted each image to remove sky background flux, and then cleaned bad pixels. Due to the extended debris disk visible in all NIRCam images, the automatic bad pixel identification routine in \texttt{spaceKLIP} did not work satisfyingly well and we used a custom bad pixel map in addition to the DO\_NOT\_USE pixels flagged by the \texttt{jwst} pipeline. This custom map was obtained by inspecting the cleaned images by eye and manually flagging bad pixels that the \texttt{jwst} pipeline had missed. While we found no additional bad pixels in the SW images, we identified and manually flagged 28 bad pixels in the LW images. The routines that were used to clean the flagged bad pixels are the same as the ones in \citet{carter2023}.

We include additional steps to mitigate the flux from the debris disk and to deal with spatial undersampling. The shortest filters in each channel are spatially undersampled, which can lead to numerical artifacts when shifting or rotating images. Before recentering and aligning the images, we first applied a Gaussian high-pass filter to subtract spatially extended and smooth flux from the debris disk. We explored a variety of filter sizes ranging from standard deviations of 1--9~pixels, as well as no high-pass filtering at all. To address the undersampling, we then blurred all images with a Gaussian kernel with a full width half maximum (FWHM) of
\begin{eqnarray}
    \rm{FWHM}_{\rm{filter}} &= \sqrt{\rm{FWHM}_{\rm{desired}}^2 - \rm{FWHM}_{\rm{current}}^2} \\
    &= \sqrt{(2.3 \cdot 1.5)^2 - \left(\frac{\lambda_{\rm{min}}}{D \cdot s}\right)^2}.
\end{eqnarray}
Here, $\lambda_{\rm{min}}$ is the minimum wavelength of the filter bandpass, $D$ is the effective telescope diameter (5.2~m for NIRCam coronagraphy due to the undersized Lyot stops), $s$ is the detector pixel scale, and the desired FWHM is $2.3 \cdot 1.5 = 3.45$~pixels\footnote{Discrete data is Nyquist sampled if the pixel scale is $\lambda_{\rm{min}} / (2.3D)$ and we added another factor of 1.5 as a numerical margin \citep[e.g.,][]{pawley2010}.}. The blurring helps to avoid ``Fourier ripples'' when numerically shifting or rotating undersampled images \citep[due to the Gibbs phenomenon, e.g.,][]{gottlieb1997}.

\subsubsection{Image alignment and target acquisition performance}
\label{sec:image_alignment_and_target_acuisition_performance}

\begin{figure}[!t]
    \centering
    \includegraphics[trim={1cm 0 1cm 0},clip,width=0.49\textwidth]{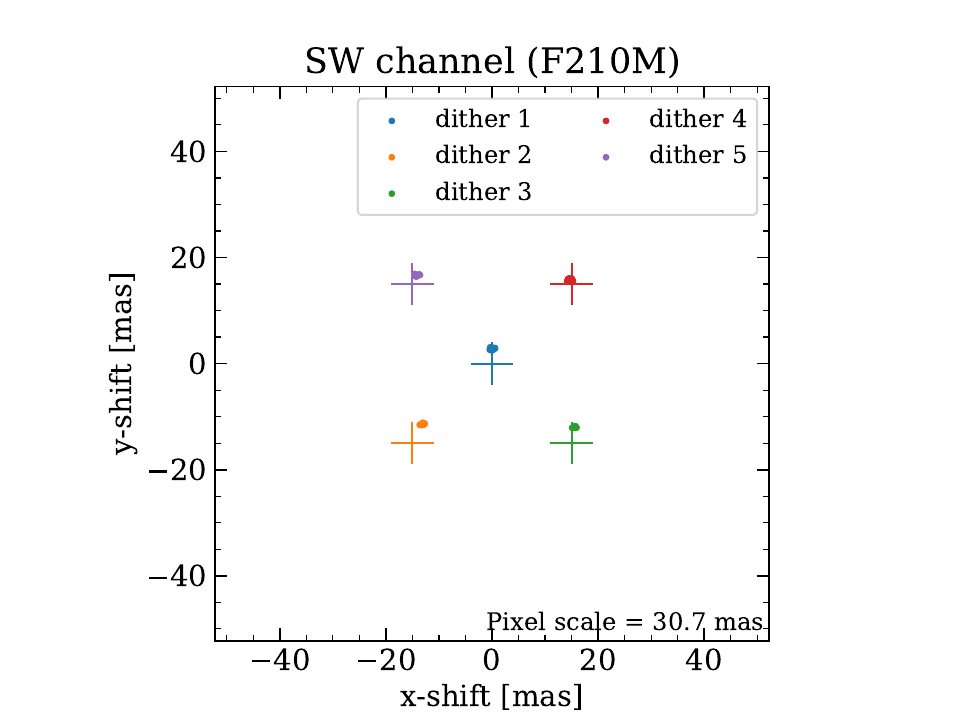}
    \includegraphics[trim={1cm 0 1cm 0},clip,width=0.49\textwidth]{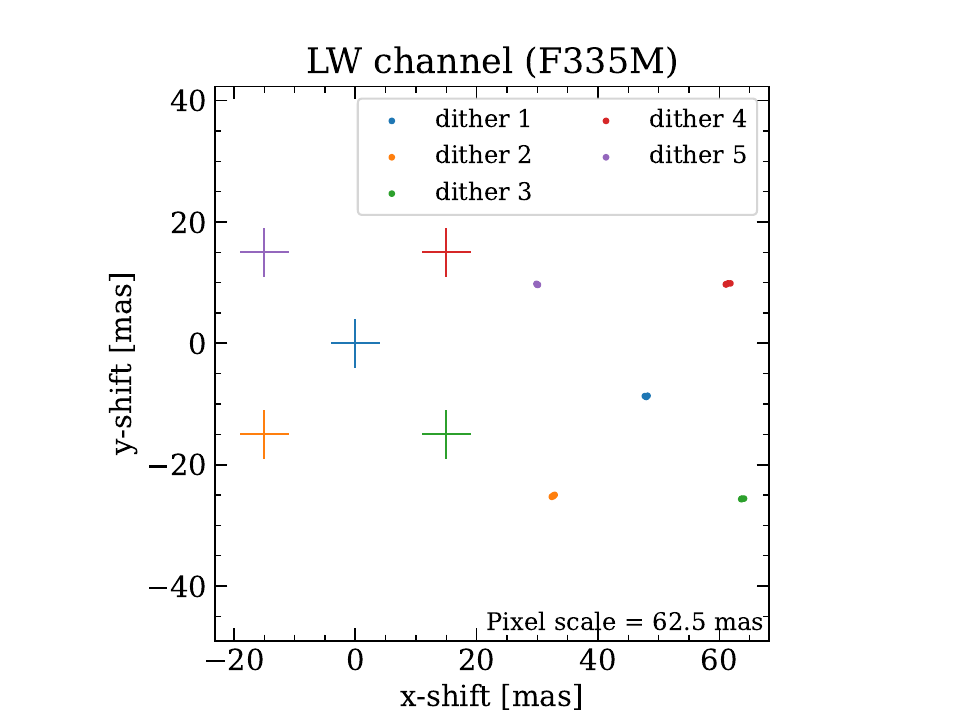}
    \caption{Reference star TA performance for the SW (top) and the LW (bottom) observations. The dots are color-coded by dither position and show the measured offsets between the reference star and the science target observation (first roll), while the crosses show the commanded and expected offsets in case of a perfect TA. The TA performance is good for the SW channel, but worse than expected for the LW channel (see Section~\ref{sec:image_alignment_and_target_acuisition_performance} for details). The plots show the F210M and F335M datasets; all other filters in the same NIRCam detector channels show similar performance.}
    \label{fig:alignment}
\end{figure}

\begin{figure*}[!t]
    \centering
    \includegraphics[trim={0 1.00cm 0 0},clip,width=0.75\textwidth]{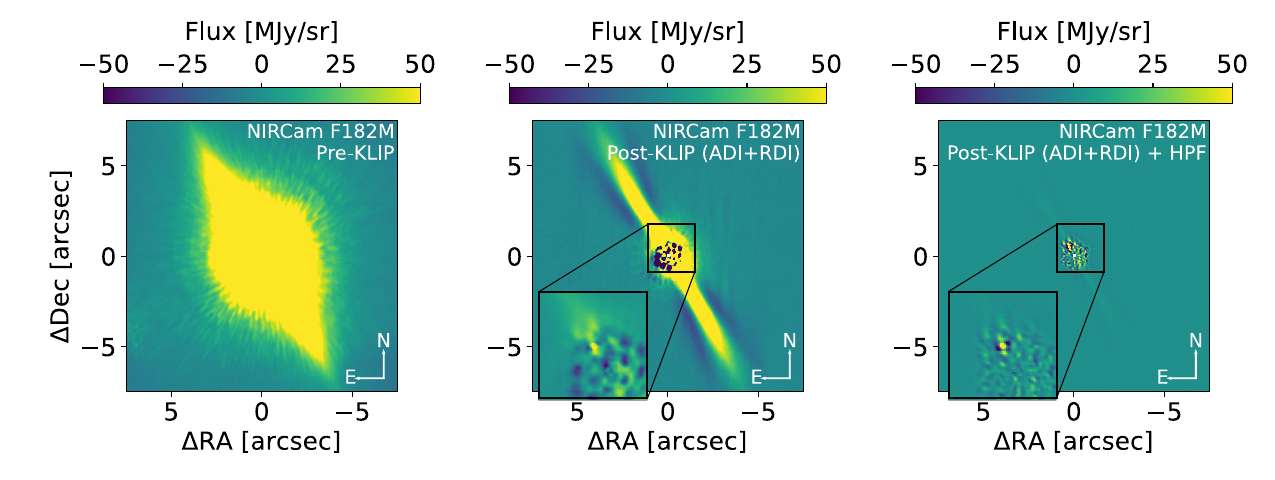}
    \includegraphics[trim={0 1.00cm 0 2.00cm},clip,width=0.75\textwidth]{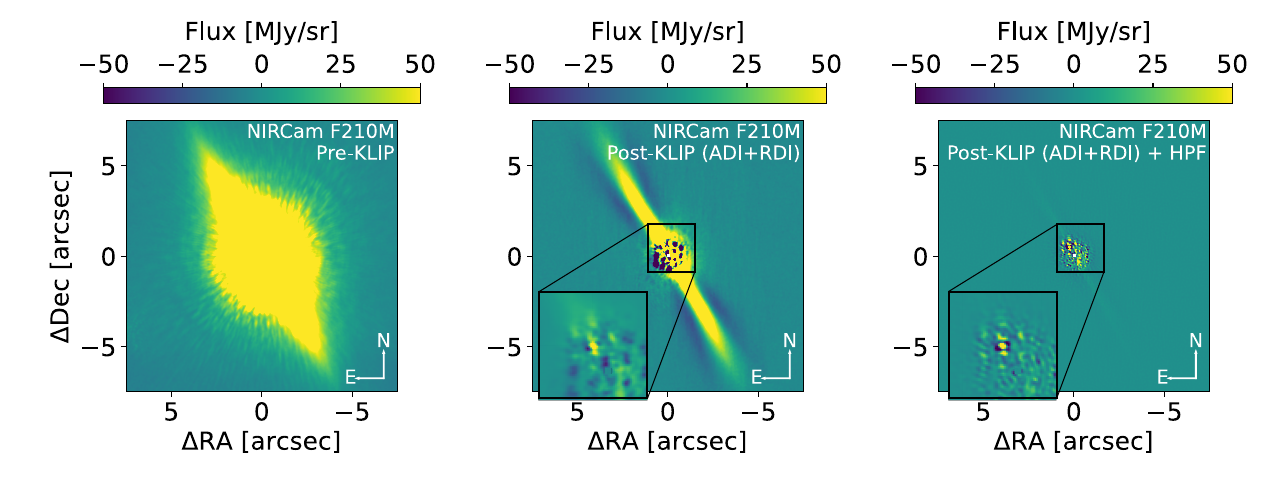}
    \includegraphics[trim={0 1.00cm 0 2.00cm},clip,width=0.75\textwidth]{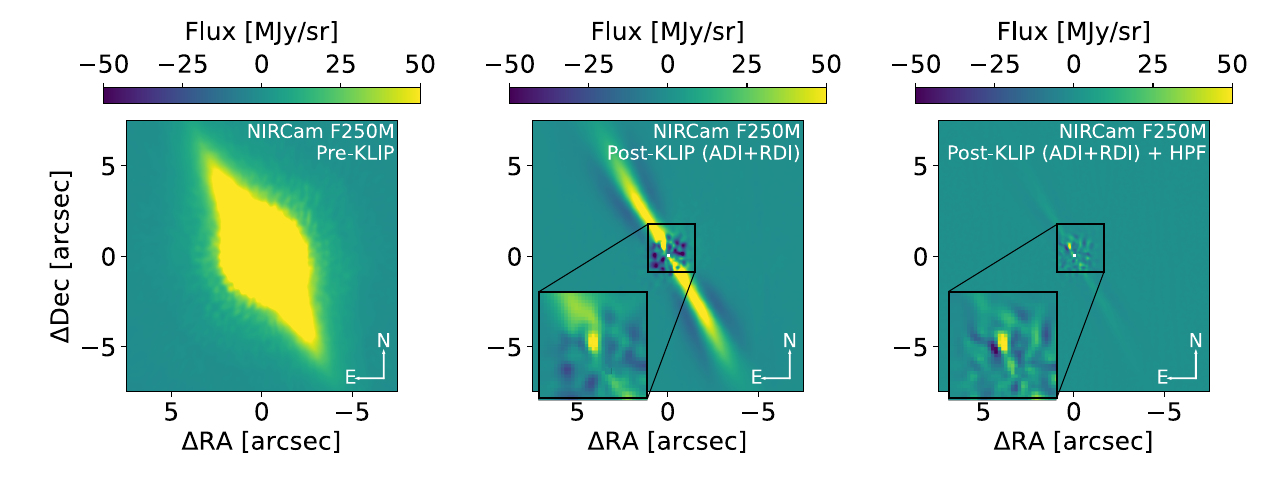}
    \includegraphics[trim={0 1.00cm 0 2.00cm},clip,width=0.75\textwidth]{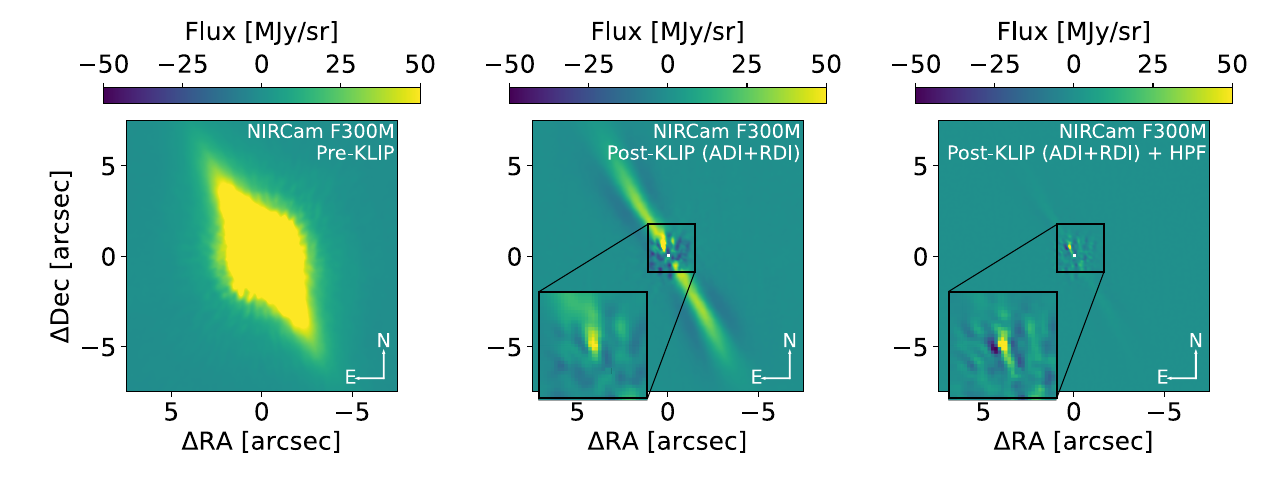}
    \includegraphics[trim={0 1.00cm 0 2.00cm},clip,width=0.75\textwidth]{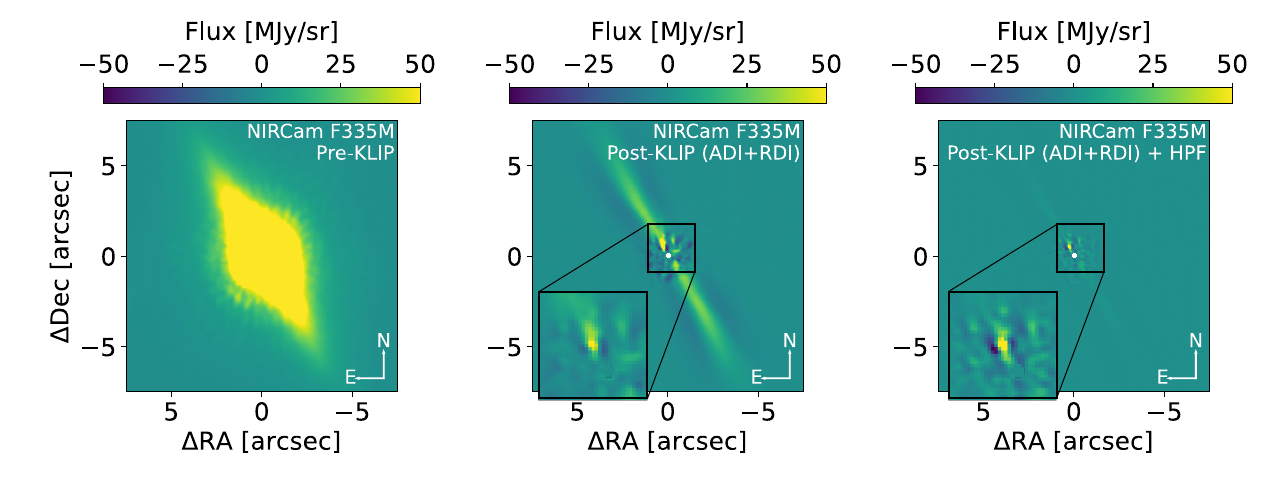}
    \includegraphics[trim={0 0 0 2.00cm},clip,width=0.75\textwidth]{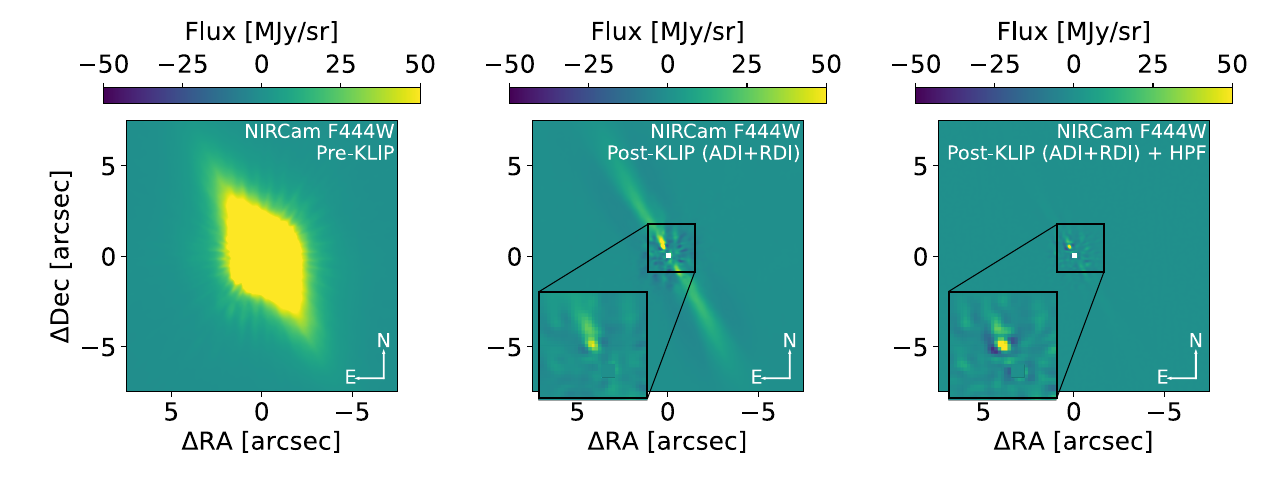}
    \caption{\emph{JWST}/NIRCam coronagraphy images of the \bpic system. The six rows show the six observed filters and the three columns show the pre-KLIP images (non-PSF-subtracted, left), the post-KLIP images (PSF-subtracted, middle), and the post-KLIP and high-pass filtered images (high-pass filter size of 3~pixels, right). The host star is located at the origin. All images are shown in the same linear color stretch. The small insets in the middle and right columns show a zoom on the giant planet \bpicb in a different color stretch that aims to highlight the planet PSF.}
    \label{fig:klip_subtraction}
\end{figure*}

Image alignment to precisely register the science images and reference PSFs is critical for achieving high-quality PSF subtractions. For each filter independently, the first science target image of each observation was recentered on the position of the target star. This is a difficult task because the star is located behind the coronagraphic mask so that its center cannot be easily determined. We followed the same procedure that was already used by \citet{greenbaum2023} to center the star based on its speckle pattern with the help of a simulated PSF computed from a contemporaneous wavefront map of the telescope using \texttt{WebbPSF\_ext}\footnote{\url{https://github.com/JarronL/webbpsf_ext}} \citep{perrin2014,girard2022}. \citet{greenbaum2023} reported a centroiding error of $\sim7$~mas for this procedure. However, given the bright debris disk around \bpic, the centroiding error is likely larger in our case. Once the first image of each observation was recentered, we then used the same image registration routine as in \citet{kammerer2022} to align all subsequent science and PSF reference target images to the first one. As discussed in \citet{rigby2023}, the pointing stability of \emph{JWST} is $\sim1$~mas, and we are able to recover a $\sim1$~mas root-mean-square (RMS) jitter between individual images. We are also able to recover the injected 5-POINT-BOX SGD for the PSF reference target and to measure the relative TA offset between the science and the PSF reference target.

Figure~\ref{fig:alignment} shows this alignment for one of the SW and one of the LW filters, revealing good TA performance with an offset of only $\sim4$~mas for the SW channel, but poor performance with an offset of $\sim50$~mas for the LW channel PSF reference observation. This is atypical, in fact we believe one of the largest-observed TA offsets in the ensemble of NIRCam coronagraphy data yet obtained. We subsequently determined that the reason for this outlier TA performance is that the SIMBAD coordinates and proper motion of \apic are not sufficiently accurate\footnote{SIMBAD currently reports the proper motion from the Hipparcos catalog, which these observations show is inaccurate over several decades. Gaia DR3 does not yet report a proper motion for this very bright star. There is a long history of \apic having discrepant proper motions between various measurements; see discussion in \citet{goldin2007}. The recent USNO Bright Star Catalog \citep{zacharias2022} reports a position and proper motion for \apic which are well consistent with its observed position in 2022 as measured with \emph{JWST}.}, so that after the initial telescope slew and fine guidance sensor acquisition, \apic was offset by $\sim0.7$~arcsec from the expected position. As a result, its PSF did not entirely fit within the NIRCam LW TA subarray. This resulted in an inaccurate centroid measurement. We note that the same systematic position offset of \apic was also seen in the NIRCam SW and MIRI TA images, but for those modes the relative size of the PSF with respect to the TA subarray is smaller so that the PSFs did still fit within the TA subarrays and the centroid measurements were accurate. We expect the poor NIRCam LW TA performance to impact the quality of the RDI PSF subtraction at small separations from the target for the LW datasets, specifically to reduce the contrast performance by a factor of $\sim3$--4 at a separation of one arcsecond with respect to what would have been achievable with a good TA performance \citep[see Figure~16 in][]{girard2022}.

After image registration, we then averaged all aligned integrations in an individual exposure (each dither position is an individual exposure) to speed up the subsequent processing with \texttt{pyKLIP}. This is beneficial for rerunning the reduction several times with different parameters later on in the paper. We note that averaging the images has a negligible impact on the KLIP reduction because the line-of-sight pointing and PSF of \emph{JWST} are extremely stable. For comparison, we also ran the \texttt{pyKLIP} processing with our baseline parameters once without averaging the images and found that $\sim99.9\%$ of the signal is contained in the first KL mode of each individual exposure which represents the noise-weighted average of the individual images in that exposure. In other words, there is negligible information loss in averaging together all integrations within an exposure.

\subsubsection{PSF subtractions (KLIP)}
\label{sec:psf_subtractions_klip}

We tested and compared several methods for PSF subtractions, including reference differential imaging \citep[RDI,][]{lafreniere2007}, angular differential imaging \citep[ADI,][]{marois2006}, and model-constrained reference differential imaging \citep[MCRDI,][]{lawson2022}. The former two were employed using the implementation of the Karhunen-Lo\`eve Image Processing (KLIP) algorithm \citep{soummer2012} in \texttt{pyKLIP} \citep{wang2015} through the \texttt{spaceKLIP} community pipeline and are described in this Section, and MCRDI is described in the following Section.

To remove post-coronagraph residual stellar speckles from the NIRCam images of \bpic, we employed ADI, RDI, and ADI+RDI techniques through \texttt{pyKLIP}. The objective of the KLIP algorithm is to project each science target image onto a covariance-weighted orthogonal basis of eigenimages of the PSF reference library, which consists of the PSF reference images of all five dither positions in the case of RDI, or the science target images from the other telescope roll in the case of ADI. The reference library constructed from the SGD observations of the PSF reference star will hence capture changes in the PSF shape as a function of the TA offset in order to provide a good representation of the science target image which is similarly affected by a TA uncertainty of typically up to $\sim20$~mas \citep{girard2022}. Then, the projection of the science target image onto the reference library is subtracted from the science target image itself, ideally leaving behind flux from off-axis companions and circumstellar structure.

For \bpic, RDI subtraction alone does not work well due to the extended debris disk contributing significant flux in all NIRCam images, and additionally the poor PSF reference star TA performance in the LW channel. The KLIP routine tends to oversubtract the residual stellar speckles because it tries to minimize the total residual flux, including the flux of the debris disk. ADI alone yields a much better speckle subtraction, but its performance at small angular separations is limited due to the small telescope roll angle of only 10~deg.

Combining ADI and RDI yields the visually best performance. The KLIP-subtracted NIRCam images are shown in Figure~\ref{fig:klip_subtraction}. We did not split the images into multiple annuli or subsections because we did not find any improvement in fake companion injection and recovery tests from doing that (see Appendix~\ref{sec:fake_companion_injection_recovery_tests}). We note that these PSF subtractions are optimized for measurements of the planet \bpicb; different choices are optimal for the study of the debris disk, as discussed in \citet{rebollido2024}.

\subsubsection{PSF subtractions (MCRDI)}
\label{sec:psf_subtractions_mcrdi}

\begin{figure*}[!t]
    \centering
    \includegraphics[trim={0 2.0cm 6.0cm 1.5cm},clip,width=0.60\textwidth]{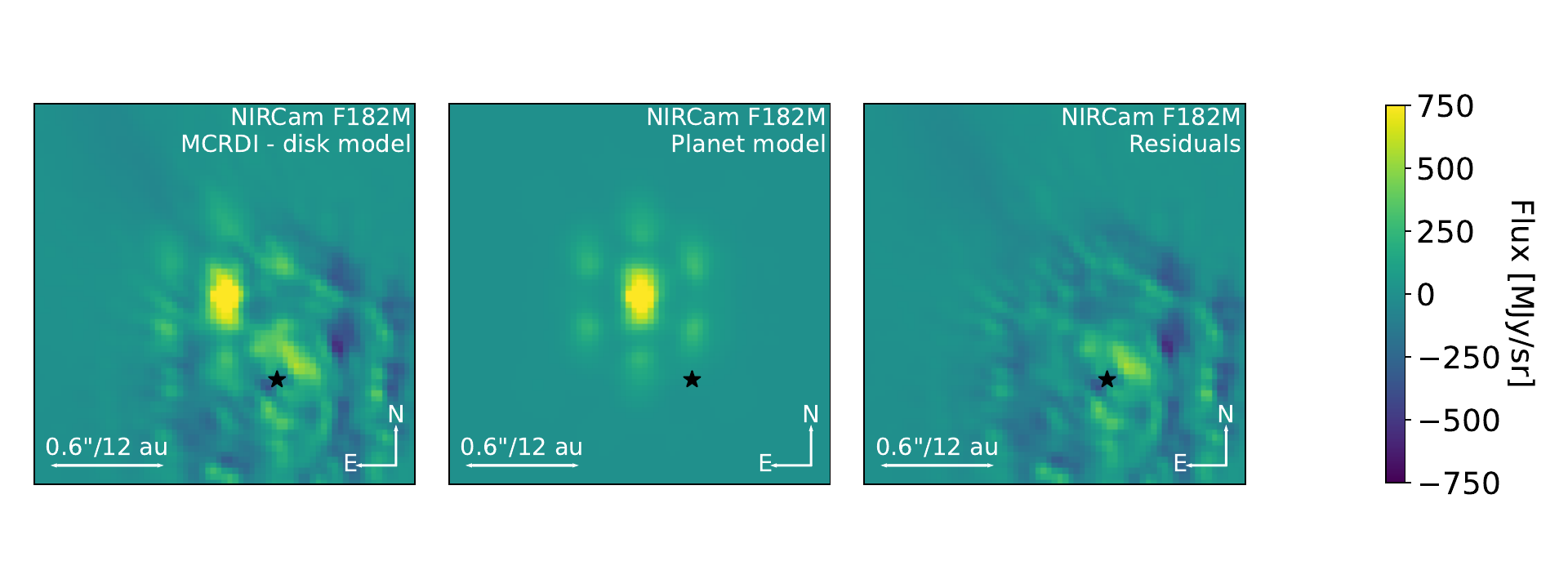}
    \includegraphics[trim={28.5cm 1.5cm 0 1.5cm},clip,width=0.08\textwidth]{figures_new/mcrdi_reduction_F182M.pdf}
    \includegraphics[trim={0 2.0cm 6.0cm 1.5cm},clip,width=0.60\textwidth]{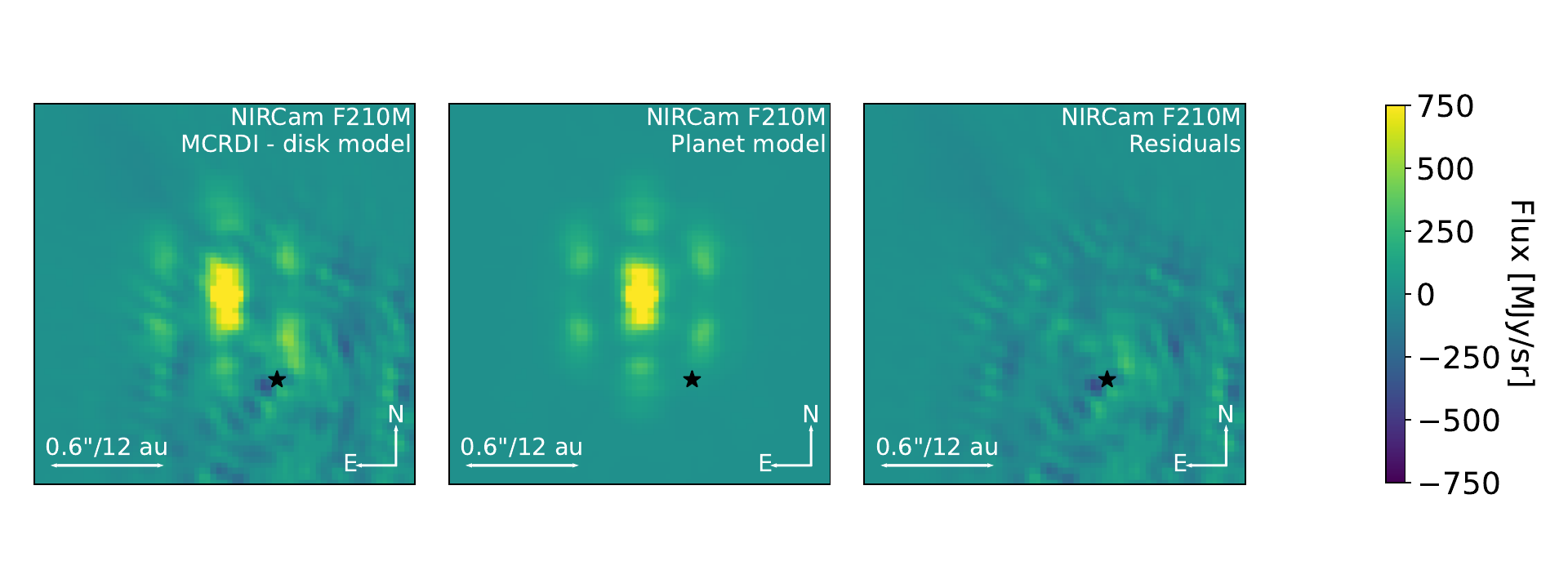}
    \includegraphics[trim={28.5cm 1.5cm 0 1.5cm},clip,width=0.08\textwidth]{figures_new/mcrdi_reduction_F210M.pdf}
    \includegraphics[trim={0 2.0cm 6.0cm 1.5cm},clip,width=0.60\textwidth]{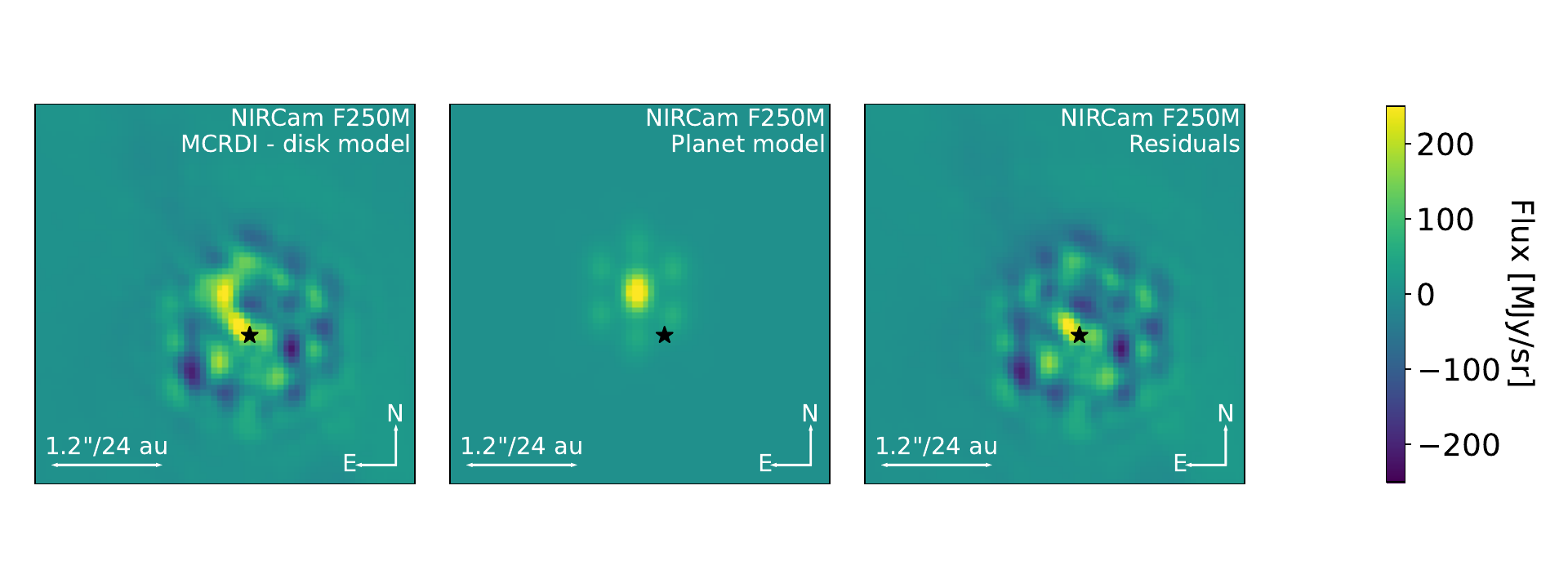}
    \includegraphics[trim={28.5cm 1.5cm 0 1.5cm},clip,width=0.08\textwidth]{figures_new/mcrdi_reduction_F250M.pdf}
    \includegraphics[trim={0 2.0cm 6.0cm 1.5cm},clip,width=0.60\textwidth]{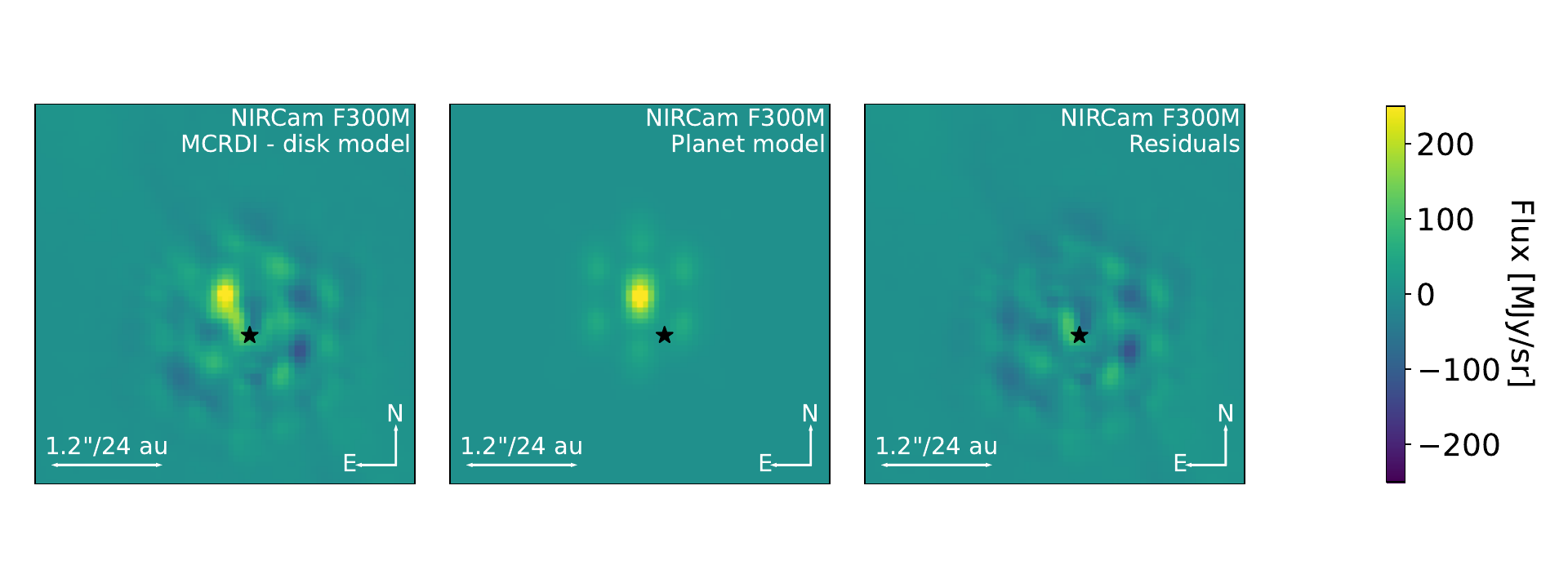}
    \includegraphics[trim={28.5cm 1.5cm 0 1.5cm},clip,width=0.08\textwidth]{figures_new/mcrdi_reduction_F300M.pdf}
    \includegraphics[trim={0 2.0cm 6.0cm 1.5cm},clip,width=0.60\textwidth]{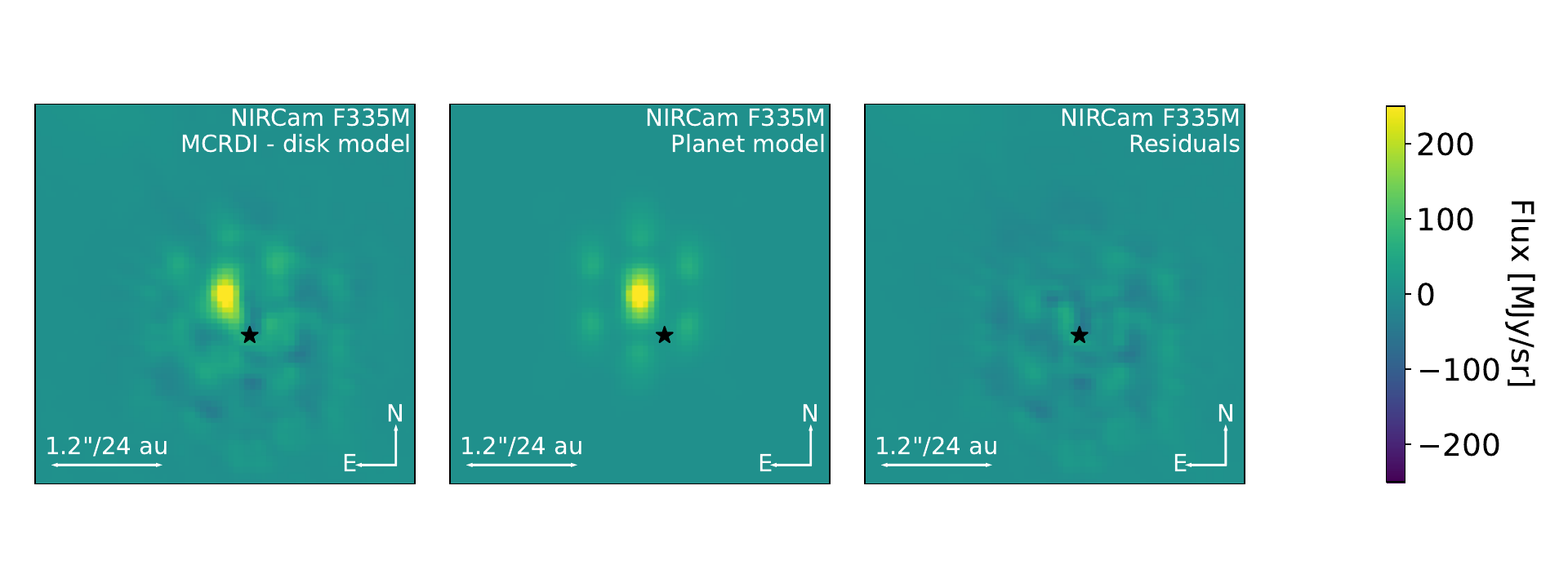}
    \includegraphics[trim={28.5cm 1.5cm 0 1.5cm},clip,width=0.08\textwidth]{figures_new/mcrdi_reduction_F335M.pdf}
    \includegraphics[trim={0 2.0cm 6.0cm 1.5cm},clip,width=0.60\textwidth]{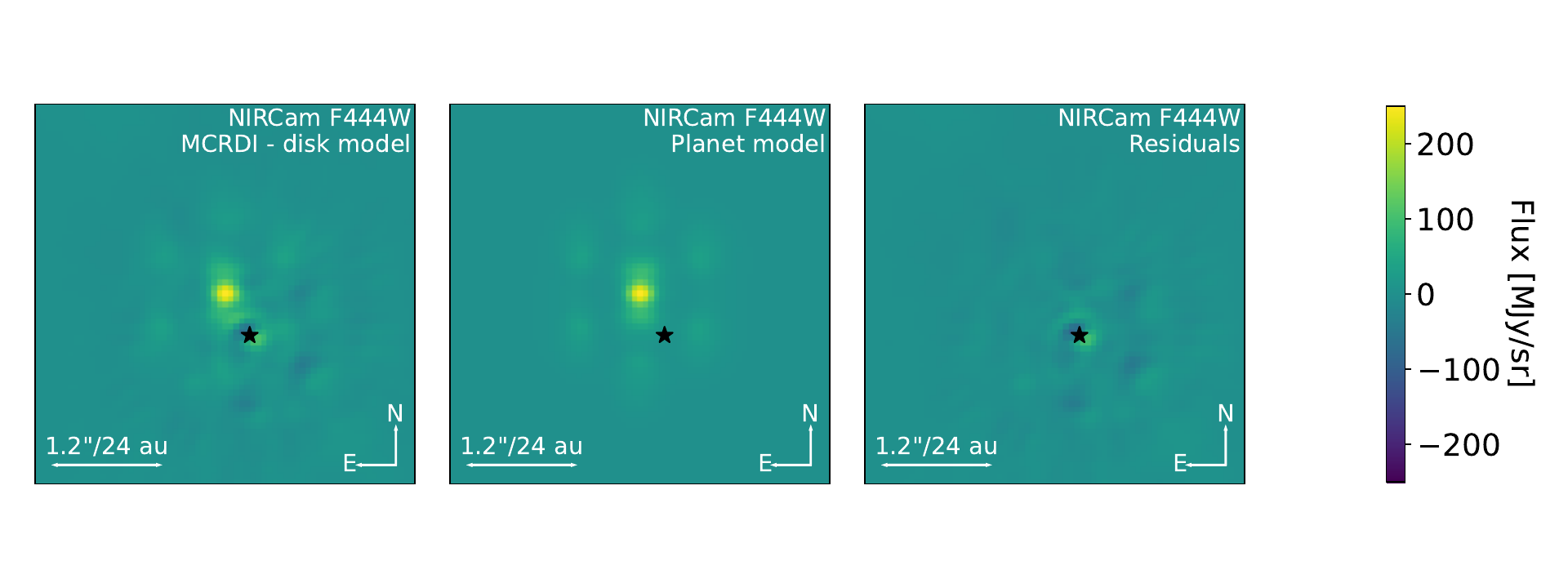}
    \includegraphics[trim={28.5cm 1.5cm 0 1.5cm},clip,width=0.08\textwidth]{figures_new/mcrdi_reduction_F444W.pdf}
    \caption{MCRDI-modeling of \bpicb. The six rows show the six observed filters and the three columns show the host star PSF- and disk model-subtracted images (left), the best fit planet PSF models (middle), and the residuals between the two (right). The position of the host star \bpic is indicated with a black star. F250M and F300M are particularly affected by spatial undersampling and the PSF subtraction has higher residuals for these filters.}
    \label{fig:mcrdi_subtraction}
\end{figure*}

In Model-Constrained RDI (MCRDI), the effects of RDI oversubtraction \citep[e.g.,][]{pueyo2016} are avoided by explicitly assuming the presence of circumstellar flux while the stellar PSF model is constructed \citep{lawson2022}. For this purpose, a synthetic model of the circumstellar scene is optimized using standard forward modeling techniques for an initial unconstrained RDI reduction. A final MCRDI reduction is then carried out using the resulting best-fit model \citep{lawson2022}. 

The MCRDI procedure predominantly follows the approach outlined in \citet{rebollido2024} but uses slightly different input data resulting from the pre-processing details described in Sections~\ref{sec:calibration_of_individual_images} and~\ref{sec:image_alignment_and_target_acuisition_performance} (e.g., the larger blurring kernel to mitigate the effects of the Gibbs phenomenon). The circumstellar model prescription effectively follows the one in \citet{rebollido2024} — assuming a circumstellar scene that is the superposition of a simple ring-like disk and a point source. We point out that this simplified disk model is only used for computing the PSF subtraction coefficients in the RDI framework, the PSF-subtracted scene images do however still comprise the disk in its full complexity. For convolution, we use a more finely spatially sampled grid of synthetic PSFs from \texttt{WebbPSF} \citep{perrin2014}, now sampling the origin along with 12 logarithmically-spaced radial positions at each of 8 linearly-spaced azimuthal positions (for 97 spatial samples in total). The finer spatial sampling has no discernible impact on the final MCRDI image, but does slightly change the disk-model-subtracted residuals within the IWA. Otherwise, the MCRDI procedure is as described in \citet{rebollido2024}.

The MCRDI-reduced NIRCam images are shown in Figure~\ref{fig:mcrdi_subtraction}, zoomed in to show just the area centered on \bpicb, after the subtraction of the disk model for each filter. Fairly good PSF subtractions are achieved for \rev{the two SW filters, but a small blob can be seen in the residuals of the F250M data towards the North-East of the host star position. We note that a similarly shaped blob can also be seen in other of the LW filters if the color stretch is optimized for the residuals instead of the planet \bpicb and the blob's brightness scales with the brightness of the other residual stellar speckles. We hence attribute this blob to the subpar TA performance for the LW observations.}

\subsection{Impact of the disk on the \bpicb photometry}
\label{sec:impact_of_the_disk_on_the_bpicb_photometry}

The debris disk around \bpic is clearly detected in all \emph{JWST}/NIRCam images and thus affects the extracted photometry of the giant planet \bpicb. The impact of this disk on the extracted planet flux in the KLIP subtractions can be complex. Residual disk flux may add on top of the planet flux, leading to an overestimation of the planet flux, or the disk may be oversubtracted and part of the planet flux may be removed, leading to an underestimation of the planet flux. Self-subtraction artifacts from running ADI on extended sources may further introduce biases in the planet flux that differ from these expectations. An important parameter that controls the impact of the disk on the extracted planet flux is the size of the high-pass filter that is used to remove (part of) the disk flux from the NIRCam images (see Section~\ref{sec:calibration_of_individual_images}). Fake companion injection and recovery tests (see Appendix~\ref{sec:fake_companion_injection_recovery_tests}) reveal that the best performance in the two SW filters is achieved using ADI alone (with a high-pass filter size of 7~pixels) whereas the best performance in the four LW filters is achieved by combining ADI and RDI (with a high-pass filter size of 3~pixels), so that we adopt these as the baseline approach for extracting planet photometry in the respective cases. We note that even with these optimized parameters, the retrieved flux in the F250M dataset is still much higher than the injected one in these tests, showing that the coarse detector sampling at $2.5~\um$ together with the extended debris disk prohibits an accurate companion flux measurement using KLIP techniques.

In the MCRDI approach, rather than having parameters for spatial filtering, instead there are additional free parameters for the disk model which are fit as part of the MCRDI process. In effect, these disk model parameters become nuisance parameters which must be solved for simultaneously, increasing the overall dimensionality of the fitting process, but providing a loosely physical approach for subtracting the disk's light around the location of the planet. As in \citet{rebollido2024}, we note that the disk model fit in MCRDI is intentionally simplified and not intended as a physically-correct model of the \bpic debris disk, but it does suffice to largely remove the disk light while avoiding biases from oversubtraction \citep{lawson2022}.

Given the substantial impact of \bpic's bright disk on the PSF subtractions, much effort and iteration went into optimizing these methods and their parameters until the two independent PSF subtraction approaches yielded consistent measurements for the photometry of \bpicb. This cross-validation was essential for establishing confidence in the removal of subtraction systematics. Potential sources of systematic errors were taken into account by inflating the error bars on the planet photometry as described in the third paragraph of Section~\ref{sec:jwst_photometry}.

\section{Results}
\label{sec:results}

\bpicb is robustly detected in all six observed filters, from which we extract photometry and astrometry measurements. We also use these data to set deep limits on the presence of additional outer planets in the system.
As expected, the inner giant planet \bpicc remains undetected behind the coronagraphic masks of NIRCam. Its angular separation at the time of these observations was only $95\pm10$~mas, resulting in a $7.3^{+0.5}_{-0.3}$~mag attenuation due to the coronagraphic masks and leading to an effective contrast of $\sim18.1$~mag or $\sim6\mathrm{e}{-8}$ in the K-band which is beyond the capabilities of NIRCam at such small separations.

\subsection{JWST photometry}
\label{sec:jwst_photometry}

\begin{figure*}[!t]
    \centering
    \includegraphics[trim={0 2.0cm 6.0cm 1.5cm},clip,width=0.60\textwidth]{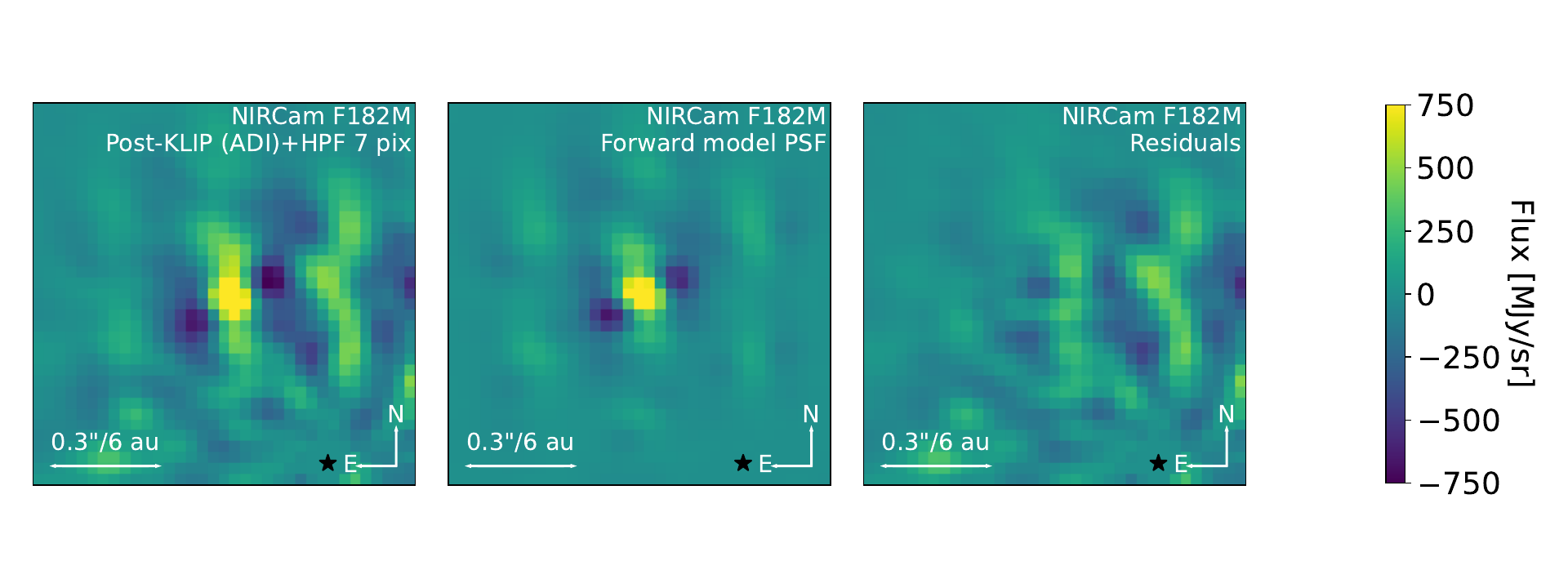}
    \includegraphics[trim={28.5cm 1.5cm 0 1.5cm},clip,width=0.08\textwidth]{figures_new/psf_fitting_F182M.pdf}
    \includegraphics[trim={0 2.0cm 6.0cm 1.5cm},clip,width=0.60\textwidth]{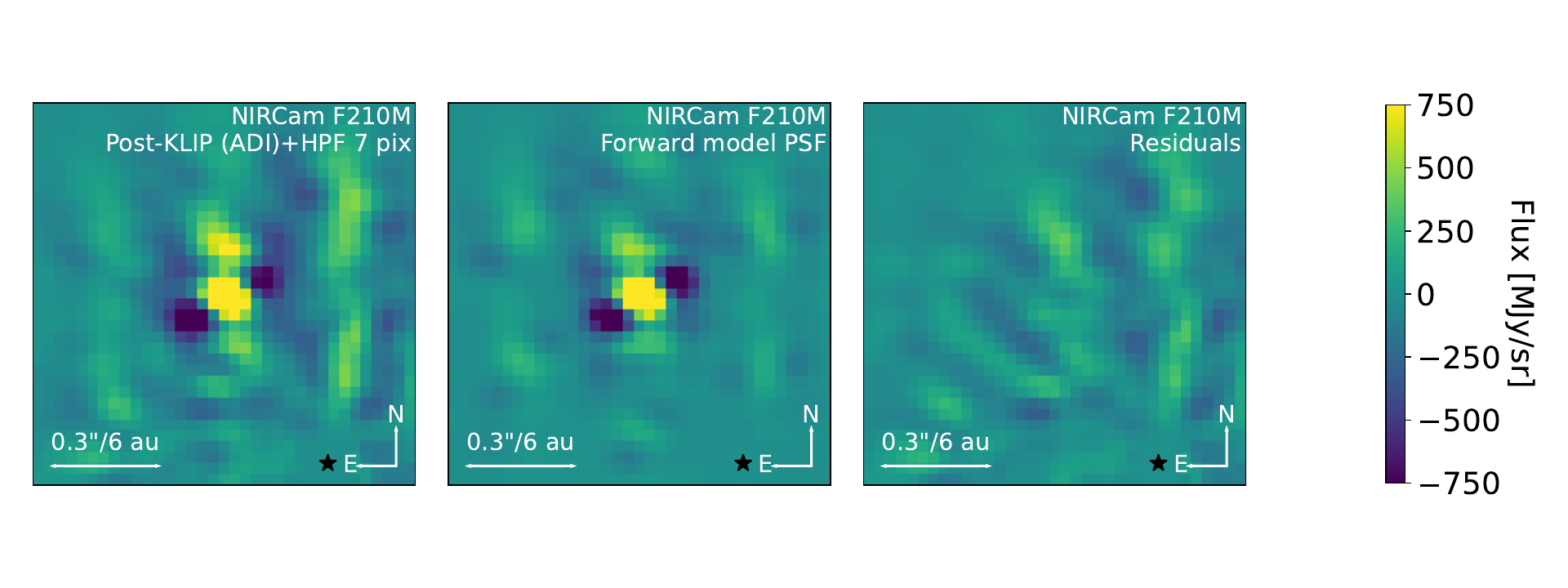}
    \includegraphics[trim={28.5cm 1.5cm 0 1.5cm},clip,width=0.08\textwidth]{figures_new/psf_fitting_F210M.pdf}
    \includegraphics[trim={0.8cm 2.2cm 5.3cm 1.5cm},clip,width=0.61\textwidth]{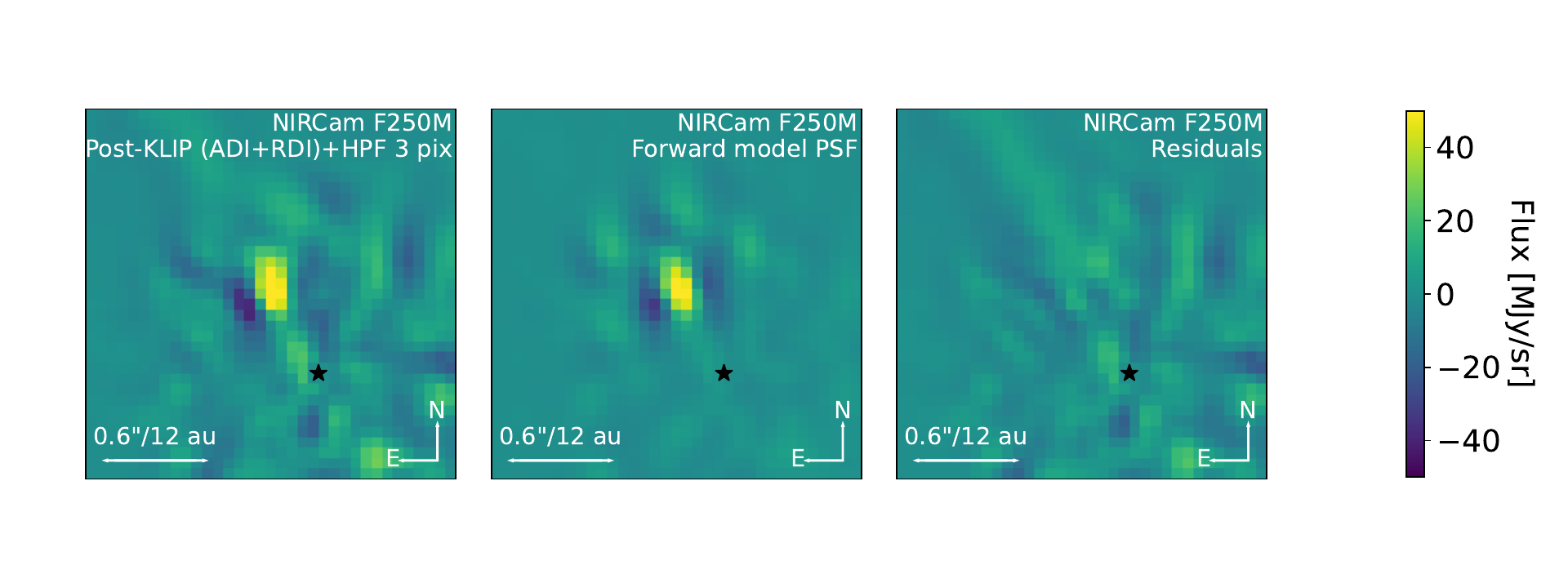}
    \includegraphics[trim={29.0cm 2.0cm 0 1.5cm},clip,width=0.08\textwidth]{figures_new/psf_fitting_F250M.pdf}
    \includegraphics[trim={0.8cm 2.2cm 5.3cm 1.5cm},clip,width=0.61\textwidth]{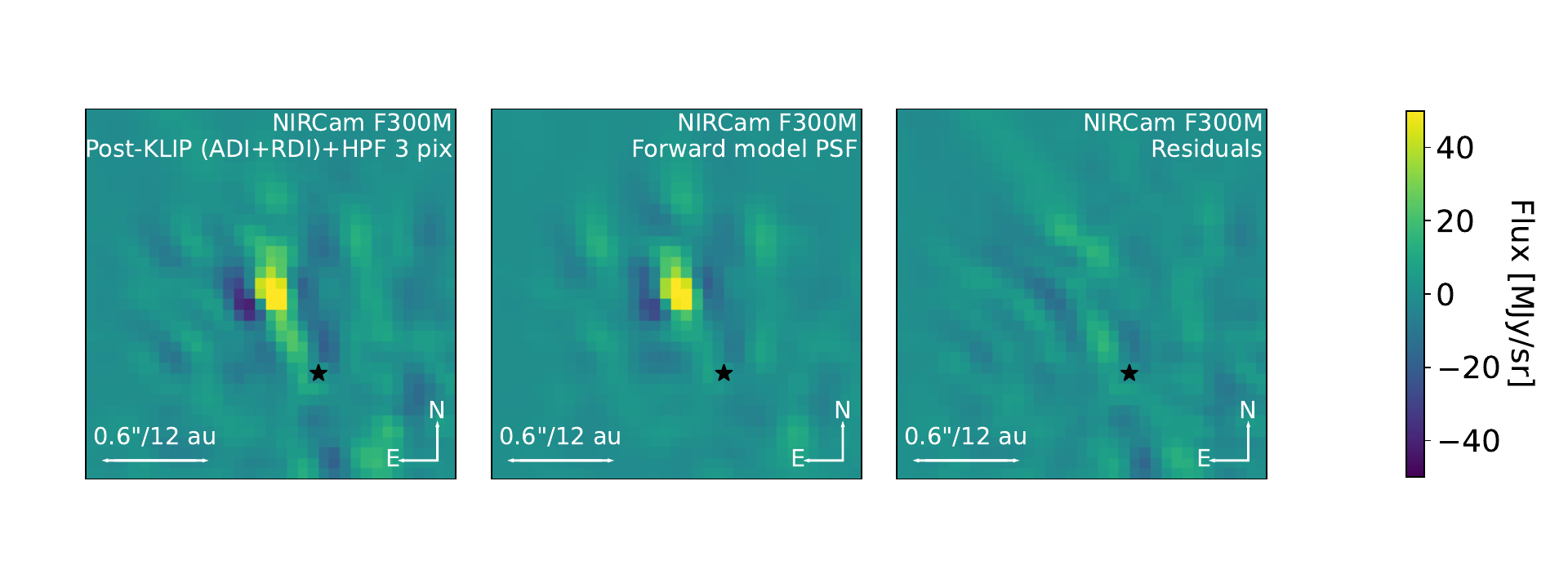}
    \includegraphics[trim={29.0cm 2.0cm 0 1.5cm},clip,width=0.08\textwidth]{figures_new/psf_fitting_F300M.pdf}
    \includegraphics[trim={0.8cm 2.2cm 5.3cm 1.5cm},clip,width=0.61\textwidth]{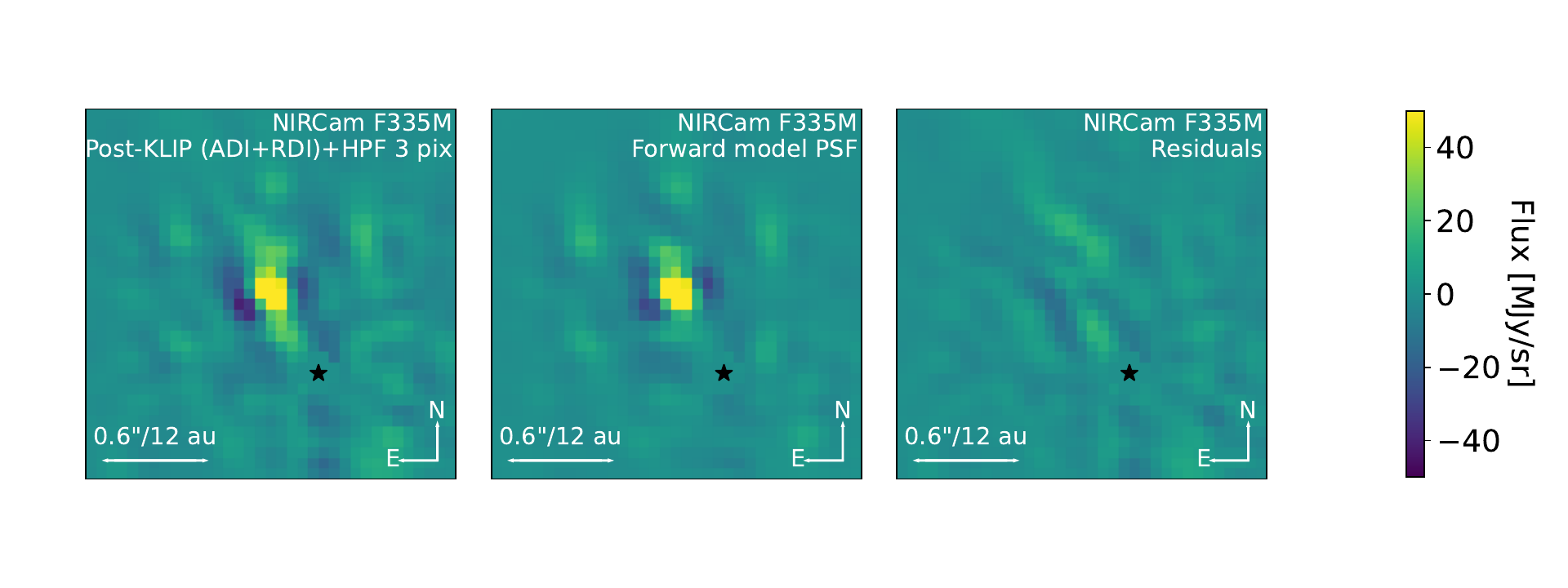}
    \includegraphics[trim={29.0cm 2.0cm 0 1.5cm},clip,width=0.08\textwidth]{figures_new/psf_fitting_F335M.pdf}
    \includegraphics[trim={0.8cm 2.2cm 5.3cm 1.5cm},clip,width=0.61\textwidth]{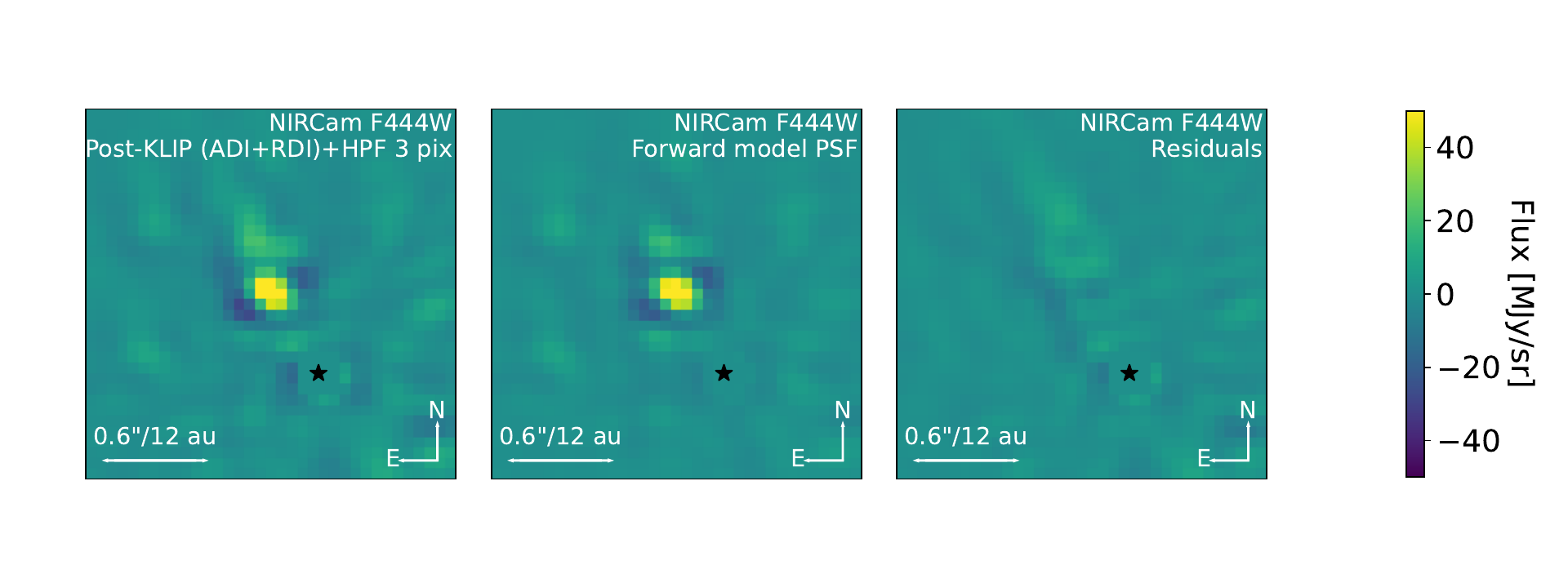}
    \includegraphics[trim={29.0cm 2.0cm 0 1.5cm},clip,width=0.08\textwidth]{figures_new/psf_fitting_F444W.pdf}
    \caption{Forward-modeling of \bpicb's PSF in the KLIP approach. The six rows show the six observed filters and the three columns show the KLIP-subtracted images (left), the best fit forward model planet PSFs (middle), and the residuals between the two (right). The position of the host star \bpic is indicated with a black star.}
    \label{fig:psf_fitting}
\end{figure*}

We extracted new photometry for the giant planet \bpicb in all six observed filters, spanning a wavelength range of $\sim1.7$--$5~\um$. This new photometry is especially interesting since L- and M-band observations from the ground are typically affected by a high thermal background from the Earth's atmosphere, leading to large uncertainties in the measured photometry \citep[e.g.,][]{stolker2020}. Our new measurements from \emph{JWST} are not affected by this issue. Furthermore, they also open up bands such as F182M and F300M which are inaccessible from the ground due to telluric absorption from water vapor in the Earth's atmosphere.

The photometry was obtained by forward-modeling the planet PSF in all of the six observed NIRCam bands and fitting it to the KLIP-subtracted images using Forward Model Matched Filtering \citep[FMMF,][]{pueyo2016}. The methods used follow the ones in \citet{kammerer2022} and \citet{carter2023}, but now using new in-flight photometric calibrations for NIRCam coronagraphy instead of the pre-flight ones. Briefly, a model PSF of \bpicb was computed at its expected distance from the coronagraphic mask center with the \texttt{WebbPSF\_ext} package using the closest in time available \emph{JWST} wavefront measurement from MAST. \rev{This PSF model accounts for the attenuation by the coronagraphic mask.} Due to different TA and therefore coronagraphic mask offsets, this needed to be done separately for each of the two telescope rolls. The model PSFs were then fed into the \texttt{pyKLIP} BKA routine \citep{wang2016} which forward-models the KLIP-processed PSF to account for companion self-subtraction (due to ADI) and KLIP algorithm throughput losses. Finally, the forward-modeled PSFs were fitted to the KLIP-subtracted images using Markov Chain Monte-Carlo (MCMC) sampling with \texttt{emcee} \citep{foreman-mackey2013}. The planet photometry (and astrometry, see next Section), including uncertainties, were obtained from the converged MCMC chains. An advantage of \emph{JWST} over ground-based imagers is that the planet photometry can directly be measured in the images since they were flux-calibrated by the \texttt{jwst} pipeline, except for the coronagraphic mask throughput which is accounted for separately in our PSF forward models. We note that this entire procedure is integrated within a single user-friendly function within \texttt{spaceKLIP}. Figure~\ref{fig:psf_fitting} shows our best fit forward model PSFs compared to the data.

\begin{figure*}[!t]
    \centering
    \includegraphics[width=0.75\textwidth]{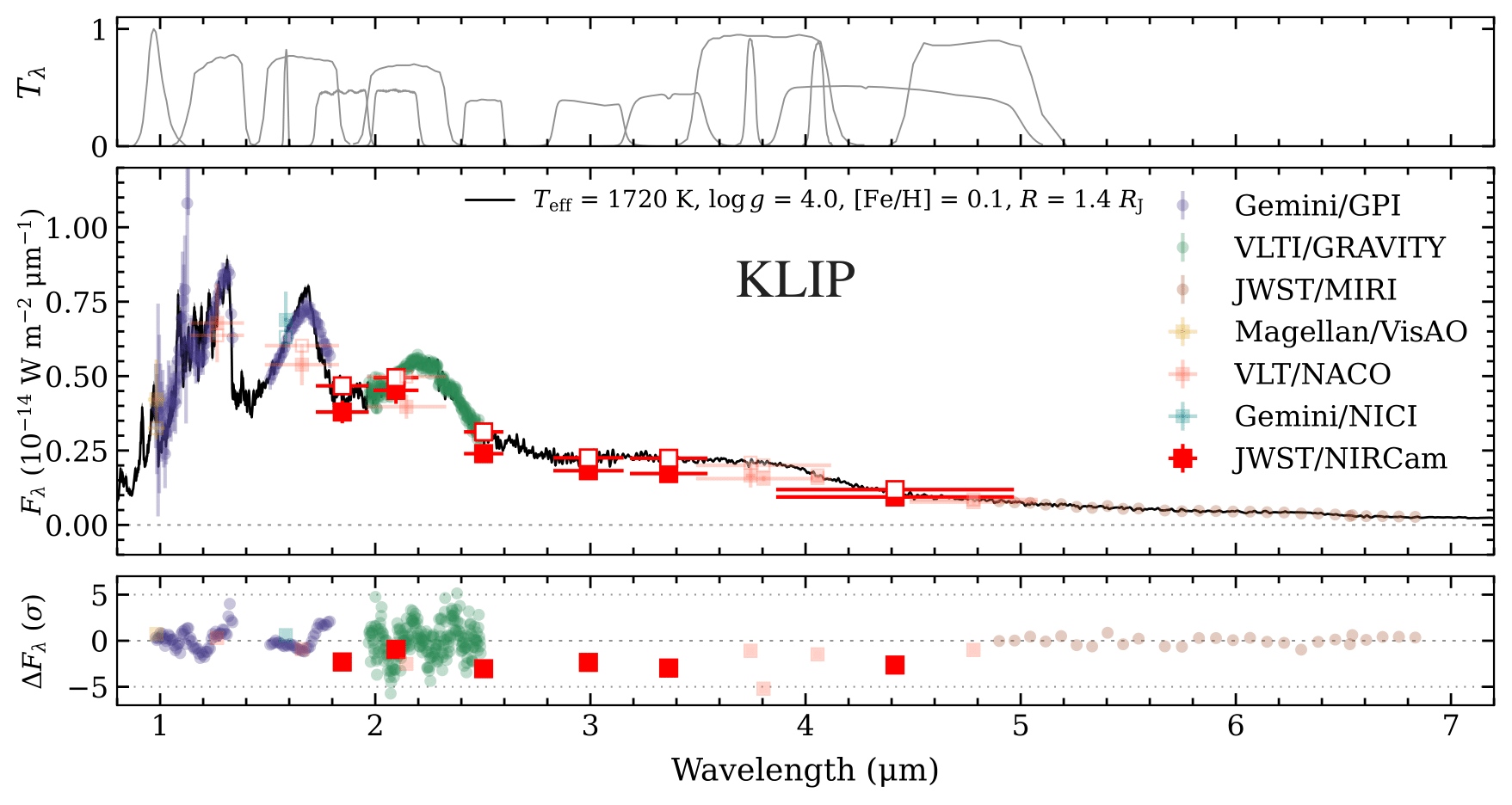}
    \includegraphics[width=0.75\textwidth]{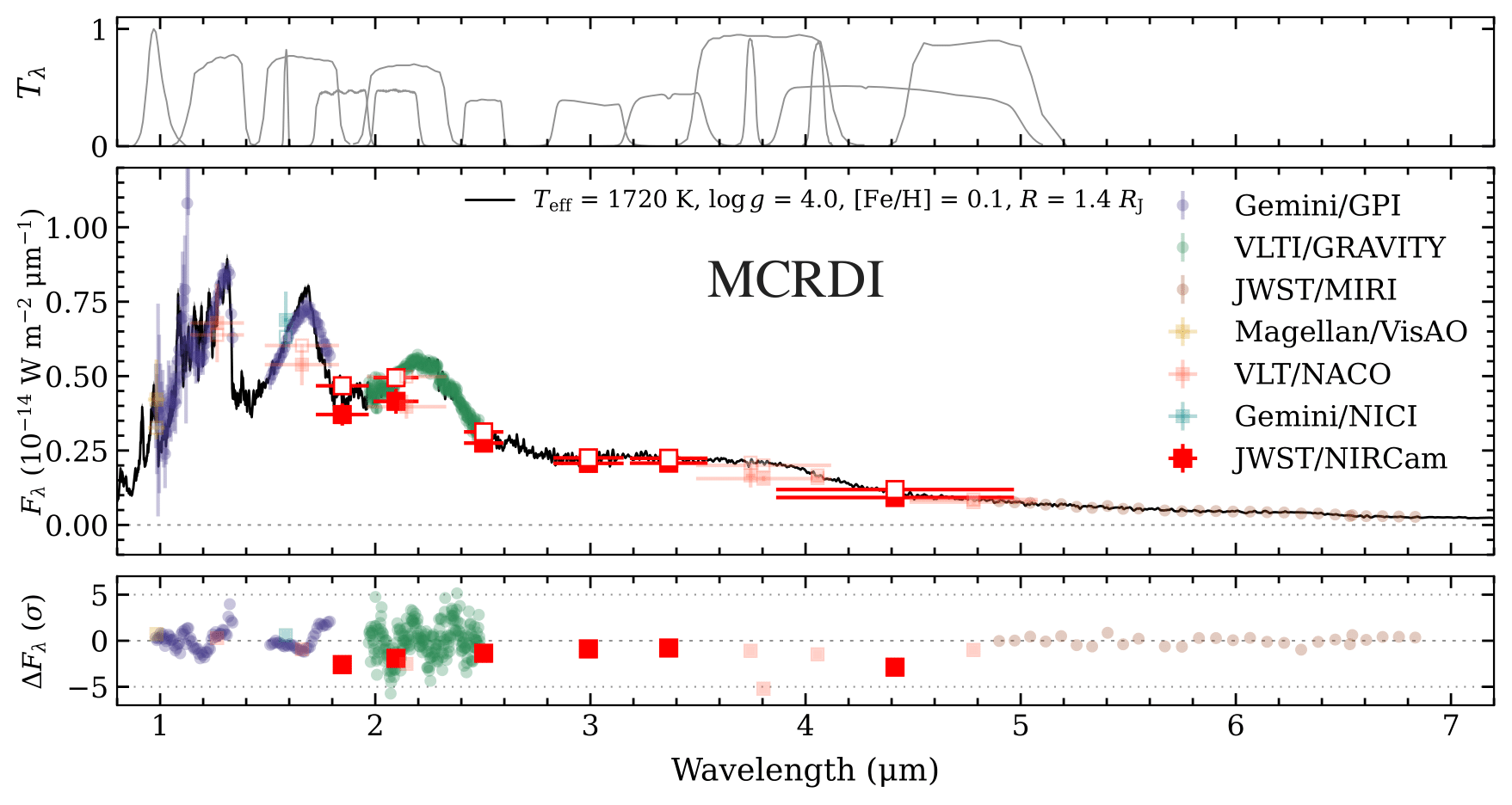}
    \caption{\emph{JWST}/NIRCam KLIP (top) and MCRDI (bottom) photometry of the giant planet \bpicb in red plotted together with other photometry and spectra and a DRIFT-PHOENIX model atmosphere of \bpicb from the literature. The filled symbols show the data and the open symbols show the values predicted by the best fitting model atmosphere. The model atmosphere was fitted to the literature data only (and not the new NIRCam data) and its parameters are printed at the top of the middle panel. The residuals between the data and the model atmosphere are shown in the bottom panel and the filter transmission curves are shown in the top panel.}
    \label{fig:photometry}
\end{figure*}

The new \bpicb NIRCam photometry is shown in Table~\ref{tab:photometry} and Figure~\ref{fig:photometry}. The KLIP photometry was obtained using the baseline extraction parameters defined in Section~\ref{sec:impact_of_the_disk_on_the_bpicb_photometry}. Besides the statistical uncertainties from the MCMC fit of the forward-modeled PSF to the data, we estimate systematic uncertainties of \rev{$\sim2\%$ from the uncertainty of the absolute flux calibration of NIRCam\footnote{\url{https://jwst-docs.stsci.edu/jwst-calibration-status/nircam-calibration-status/nircam-coronagraphy-calibration-status}}}, $\sim1\%$ from numerical inaccuracies in the forward-modeled PSFs from \texttt{WebbPSF}, and $\sim2\%$ from the uncertainty on the coronagraphic mask throughput (due to the uncertainty on the companion/mask position) that need to be added in quadrature to the statistical uncertainties. \rev{The companion flux uncertainty from the uncertainty on the companion/mask position was derived by shifting companion and mask closer together and further away from each other by their respective position uncertainties and evaluating the resulting change in flux using \texttt{WebbPSF}. The uncertainty in the \texttt{WebbPSF} models is motivated by \citet{weisz2024} who find sub-1\% photometric errors for \texttt{WebbPSF} models in regular NIRCam imaging, and we budget 1\% for this effect to be slightly conservative for the more complex optical situation of coronagraphy.} In addition, and based on the results from the fake companion injection and recovery tests (Appendix~\ref{sec:fake_companion_injection_recovery_tests}), we added another systematic error of 5\% to the KLIP photometry which accounts for the typical offset that we find between the injected and recovered fake companion flux.

Figure~\ref{fig:photometry} shows the KLIP and the MCRDI photometry together with existing photometry and spectra and a DRIFT-PHOENIX model atmosphere of \bpicb from the literature. The model atmosphere was fitted to the literature data only (and not to the new NIRCam data) using the \texttt{species}\footnote{\url{https://github.com/tomasstolker/species}} toolkit \citep{stolker2020}. The purpose of this Figure is to show whether the new NIRCam photometry agrees with the existing literature data and qualitatively assess whether the NIRCam data points at wavelengths inaccessible from the ground are broadly consistent with models used to interpret the ground-based data; model atmosphere fitting including the new NIRCam photometry will be more rigorously conducted in Section~\ref{sec:the_atmosphere_of_bpicb}. Figure~\ref{fig:photometry} shows that the new NIRCam photometry agrees well with a DRIFT-PHOENIX model atmosphere fitted to previously existing literature data of \bpicb. Both the KLIP and the MCRDI data points fall within $3\sigma$ from the prediction of the model atmosphere. 

Notably, all NIRCam data points fall below the prediction of the model atmosphere. In the L- and M-band, this makes the new NIRCam photometry consistent with the existing VLT/NACO photometry which also falls slightly below the prediction of the model atmosphere. In the K-band, however, there is a $\sim10$--$20\%$ discrepancy between the new NIRCam photometry and the existing VLTI/GRAVITY spectrum. We note that the blurring, although being accounted for in the forward modeling, might be responsible for blending some of the planet flux with the disk flux and leading to an underestimation of the planet flux. The disk being brighter at shorter wavelengths makes this especially problematic in the K-band. We further note that the fake companion injection and recovery tests (Figure~\ref{fig:injection_recovery}) show that the retrieved planet flux declines with increasing high-pass filter size. In these tests, a rather large FWHM was ideal to retrieve the injected flux for the fake companion, but it could well be that for the real planet \bpicb, a smaller FWHM might be ideal so that we are underestimating its flux. This is extremely difficult to quantify though, since the PSF- and disk-subtraction residuals are not exactly the same on the North-East and the South-West side of the debris disk. Despite these uncertainties, it is worth mentioning that our NIRCam photometry is in good agreement with the NACO data point at $\sim2.1~\micron$ though. Besides issues with the data reduction, astrophysical variability might also play a role for the measured discrepancies. While variability in the planet itself is expected to be \rev{below $\sim5\%$ based on the observed variability of low-mass brown dwarfs and planetary-mass objects \citep{metchev2015,vos2020,vos2022}}, the findings from \citet{rebollido2024} suggest that the debris disk around \bpic is a highly dynamic environment. Moreover, Worthen et al. (in prep.) find significant variability between their MIRI/MRS and archival \emph{Spitzer} spectra \citep{lu2022} of the debris disk supporting this hypothesis. \rev{Nevertheless, the impact that this disk variability would have on the flux of the planet \bpicb is expected to be only a fraction of the disk variability itself and hence smaller than the observed discrepancy between GRAVITY and NIRCam, NACO, and GPI (see also Section~\ref{sec:the_atmosphere_of_bpicb}), so that systematics in the flux calibration remain the more likely hypothesis.}


Due to the complications of the debris disk affecting the planet flux measured in the KLIP-subtracted images and the advantages of using MCRDI in such cases as discussed in Section~\ref{sec:impact_of_the_disk_on_the_bpicb_photometry}, we decided to use the planet flux measurements from the MCRDI reductions as our baseline for the atmospheric characterization of \bpicb in Section~\ref{sec:discussion}.

\begin{table*}[!t]
    \centering
    \begin{tabular}{c c c c c c c}
        Filter & Flux KLIP [Jy] & $\Delta\rm{mag}$ KLIP & Flux MCRDI [Jy] & $\Delta\rm{mag}$ MCRDI & $\rm{TP}_{\rm{MSK}}$ & $\rm{TP}_{\rm{COM}}$ \\
        \hline
        F182M & $4.31\pm0.43\mathrm{e}{-3}$ & $9.67\pm0.04$ & $4.23\pm0.42\mathrm{e}{-3}$ & $9.69\pm0.04$ & 0.838 & 0.663 \\
        F210M & $6.62\pm0.66\mathrm{e}{-3}$ & $8.99\pm0.02$ & $6.10\pm0.61\mathrm{e}{-3}$ & $9.08\pm0.02$ & 0.832 & 0.962 \\
        F250M & $5.00\pm0.50\mathrm{e}{-3}$ & $8.96\pm0.04$ & $5.75\pm0.58\mathrm{e}{-3}$ & $8.81\pm0.04$ & 0.395 & 0.963 \\
        F300M & $5.43\pm0.54\mathrm{e}{-3}$ & $8.53\pm0.03$ & $6.19\pm0.62\mathrm{e}{-3}$ & $8.40\pm0.03$ & 0.390 & 0.898 \\
        F335M & $6.51\pm0.65\mathrm{e}{-3}$ & $8.11\pm0.02$ & $7.85\pm0.79\mathrm{e}{-3}$ & $7.91\pm0.02$ & 0.384 & 0.936 \\
        F444W & $6.08\pm0.61\mathrm{e}{-3}$ & $7.66\pm0.02$ & $6.06\pm0.61\mathrm{e}{-3}$ & $7.68\pm0.02$ & 0.383 & 0.918 \\
    \end{tabular}
    \caption{\emph{JWST}/NIRCam photometry of the giant planet \bpicb. While the flux values were directly measured from the flux-calibrated NIRCam images, the contrast ($\Delta\rm{mag}$) was computed relative to a stellar model atmosphere of \bpic and thus contains an additional systematic uncertainty. $\rm{TP}_{\rm{MSK}}$ and $\rm{TP}_{\rm{COM}}$ denote the throughput of the coronagraphic mask at the position of the planet and the throughput of the coronagraphic mask substrate, respectively.}
    \label{tab:photometry}
\end{table*}



\subsection{JWST astrometry}
\label{sec:jwst_astrometry}

\begin{figure*}[!t]
    \centering
    \includegraphics[width=0.75\textwidth]{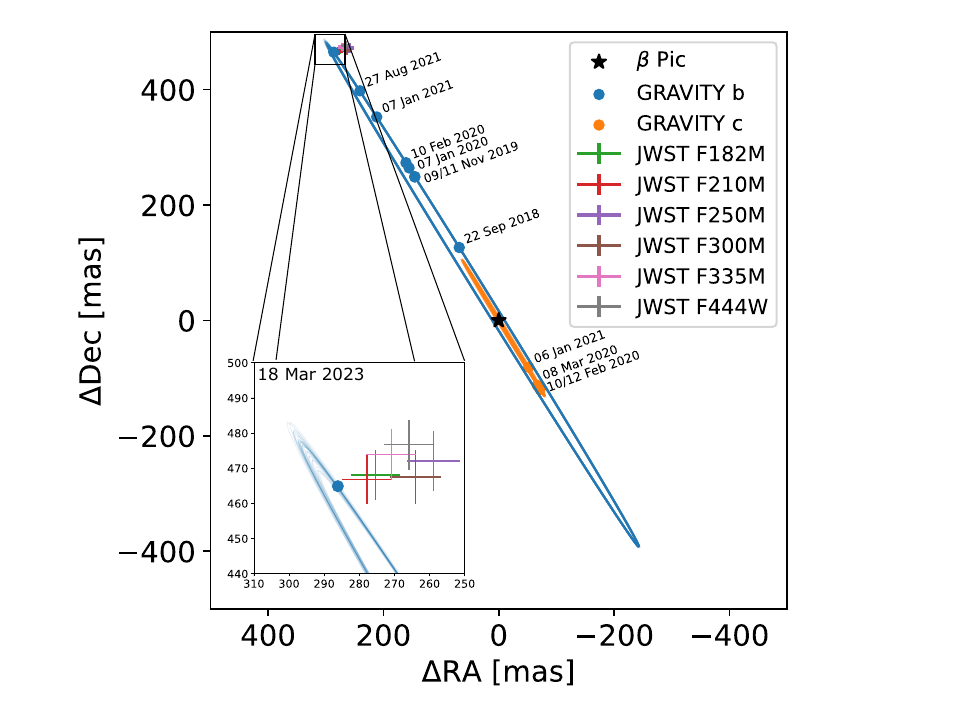}
    \caption{\emph{JWST}/NIRCam astrometry of the giant planet \bpicb plotted together with the GRAVITY data points and the inferred orbits of both \bpic planets from \citet{lacour2021}. The blue point in the inlet shows the expected position of \bpicb at the time of the \emph{JWST} observations. Due to the large (systematic) uncertainties if compared to the GRAVITY measurements, the NIRCam astrometry does not provide improved constraints on the orbital parameters of \bpicb.}
    \label{fig:orbit}
\end{figure*}

From the KLIP-subtracted \emph{JWST}/NIRCam images of the \bpic system, we also extracted relative astrometry of the well-known outer giant planet \bpicb. This astrometry is shown in Table~\ref{tab:astrometry} and Figure~\ref{fig:orbit}. It was obtained using KLIP ADI+RDI and a high-pass filter size of 3~pixels for all observed NIRCam bands\footnote{Given the large systematic uncertainties on the NIRCam astrometry, the exact choice of the extraction parameters is irrelevant here.}. \rev{While the photometry of \bpicb was taken from the MCRDI reduction, we found that the planet astrometry from MCRDI is rather sensitive to the exact choice of disk model parameters (e.g., one-component vs two-component disk model). On the other hand, the KLIP astrometry is much more consistent for different high-pass filter size choices, so that we decided to use this more robust measurement.}

In the NIRCam images, \bpicb is detected close to its maximum orbital elongation on the North-East side of the disk. The astrometric measurements from NIRCam are not yet competitive with high-contrast imagers from the ground, mainly due to the difficulties in finding the position of the star behind the coronagraphic mask and the missing distortion correction in \texttt{spaceKLIP}. Together, they lead to systematic uncertainties on the order of $\sim10$~mas in the NIRCam astrometry. We further expect that the numerous residual stellar speckles caused by the poor reference star TA performance in the LW channel negatively affect the precision of the NIRCam astrometry, so that its systematic offset with respect to the GRAVITY orbit shown in Figure~\ref{fig:orbit} is not a surprise. This expectation is supported by the LW channel astrometry generally agreeing worse with the GRAVITY orbit than the SW channel astrometry. Given these systematic errors and the precision of the existing GRAVITY astrometry of only $\sim0.1$--0.3~mas \citep{lacour2021}, we did not attempt any orbital fits for the \bpic system with our new NIRCam measurements. Instead, we only show the orbits of both planets inferred by \citet{lacour2021} from the GRAVITY data and overplot the new \bpicb NIRCam astrometry to demonstrate that we indeed detected the planet at its expected location. Improving the NIRCam astrometry and making it competitive with ground-based imagers is left for future work.

\begin{table}[!t]
    \centering
    \begin{tabular}{c c c c c}
        Filter & $\Delta\rm{RA}$ [mas] & $\Delta\rm{Dec}$ [mas] & $\rho$ [mas] & PA [deg] \\
        \hline
        F182M & $275\pm7$ & $468\pm7$ & $543\pm7$ & $30.5\pm0.7$ \\
        F210M & $278\pm7$ & $467\pm7$ & $543\pm7$ & $30.8\pm0.7$ \\
        F250M & $259\pm7$ & $472\pm8$ & $538\pm8$ & $28.7\pm0.8$ \\
        F300M & $264\pm7$ & $468\pm8$ & $537\pm8$ & $29.4\pm0.8$ \\
        F335M & $271\pm7$ & $474\pm7$ & $546\pm7$ & $29.7\pm0.8$ \\
        F444W & $266\pm7$ & $477\pm7$ & $546\pm7$ & $29.1\pm0.7$ \\
    \end{tabular}
    \caption{\emph{JWST}/NIRCam astrometry of the giant planet \bpicb as observed on 18 March 2023 (MJD = 60021). The reported uncertainties include systematic errors introduced by the limited precision with which the star's position behind the coronagraphic mask can be inferred. We report results separately for each filter here for completeness, but we note that all filters are consistent within the uncertainties with one another.}
    \label{tab:astrometry}
\end{table}

\subsection{Limits for other companions in the \bpic system}
\label{sec:limits_for_other_companions_in_the_bpic_system}

\emph{JWST}/NIRCam coronagraphy has demonstrated exquisite sensitivity to new companions at wide separations from the host star, down to the sub-Jupiter mass regime in the wide-band filters \citep[e.g.,][]{carter2023}. Here, we use the new NIRCam data of \bpic to search for previously undiscovered companions.

\begin{figure*}[!t]
    \centering
    \includegraphics[trim={1.5cm 0 18.0cm 0},clip,width=0.49\textwidth]{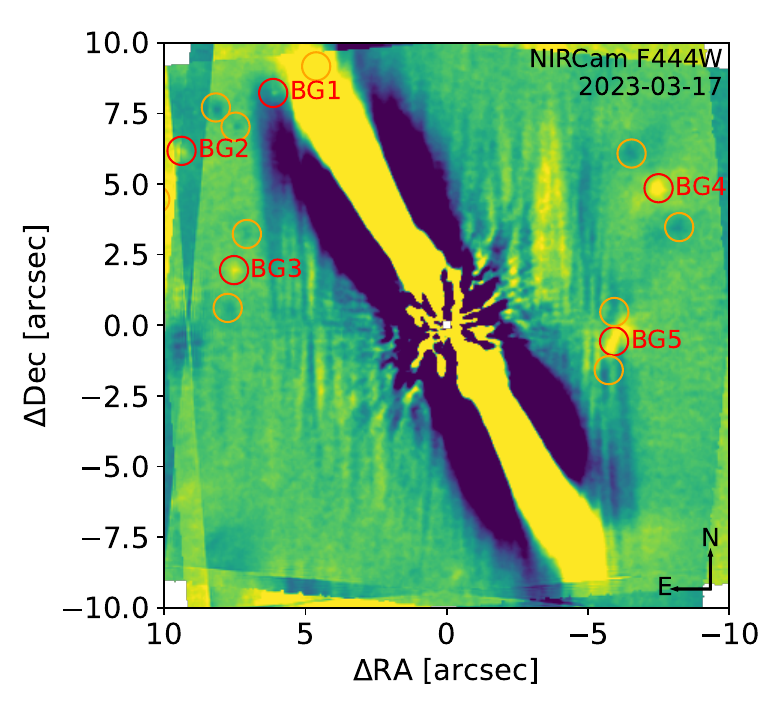}
    \includegraphics[trim={18.5cm 0 1.0cm 0},clip,width=0.49\textwidth]{figures_new/scene.pdf}
    \caption{PSF-subtracted \emph{JWST}/NIRCam (left) and MIRI (right) images of the \bpic system (arbitrary color stretch). Five background sources were identified based on their NIRCam colors or morphology whose positions are highlighted in both images with red circles. The orange circles in the NIRCam image show the source positions rotated by the telescope roll angle, where negative ADI residuals are visible. BG4 appears to be the only source which is also visible in the MIRI image. The MIRI image was adapted from \citet{rebollido2024}.}
    \label{fig:scene}
\end{figure*}

By visually inspecting the KLIP-subtracted images, we identify five other sources at separations $>5$~arcsec from \bpic in addition to the known outer giant planet \bpicb (see Table~\ref{tab:background_sources} and Figure~\ref{fig:scene}). We note again that NIRCam coronagraphy does not achieve the IWA required to detect the inner known giant planet \bpicc at a separation of $\sim100$~mas at the time of our observations. With the exception of one source (BG4), the additional sources are only detected in the NIRCam LW channel, and are visible at a low number of KL modes ($\leq4$) in all four LW filters. BG4 is also visible in the MIRI F1550C four quadrant phase mask data from \citet{rebollido2024}. Three of the five sources (BG1, BG2, BG3) are located to the East of the edge-on debris disk and the other two sources (BG4, BG5) are located to the West of the disk. The two sources to the West are clearly resolved and appear diffuse, so that we classify them as background galaxies. Two of the three sources to the East (BG2, BG3) also appear resolved so that we count four background galaxies in total. Only one of the sources to the East (BG1) appears point-like and unresolved. Its approximate position with respect to \bpic is $\Delta\rm{RA} \approx 6''$ and $\Delta\rm{Dec} \approx 8''$. Approximate astrometry of all five sources can be found in Table~\ref{tab:background_sources}.

\begin{table}[!t]
    \centering
    \begin{tabular}{c c c c}
        Source & $\Delta$RA [arcsec] & $\Delta$Dec [arcsec] & F444W flux [Jy] \\
        \hline
        BG1 & 6.057 & 8.182 & $2.2\mathrm{e}{-6}$ \\
        BG2 & 9.387 & 6.174 & $2.0\mathrm{e}{-6}$ \\
        BG3 & 7.529 & 1.953 & $2.1\mathrm{e}{-6}$ \\
        BG4 & -7.497 & 4.851 & $3.5\mathrm{e}{-6}$ \\
        BG5 & -5.922 & -0.567 & $4.6\mathrm{e}{-6}$ \\
    \end{tabular}
    \caption{Astrometry and photometry of the detected background sources. Only the values for BG1 are based on an accurate fit with a model PSF. The four other sources appear extended and their astrometric positions are the approximate centroids.}
    \label{tab:background_sources}
\end{table}

\begin{figure*}[!t]
    \centering
    \includegraphics[width=0.80\textwidth]{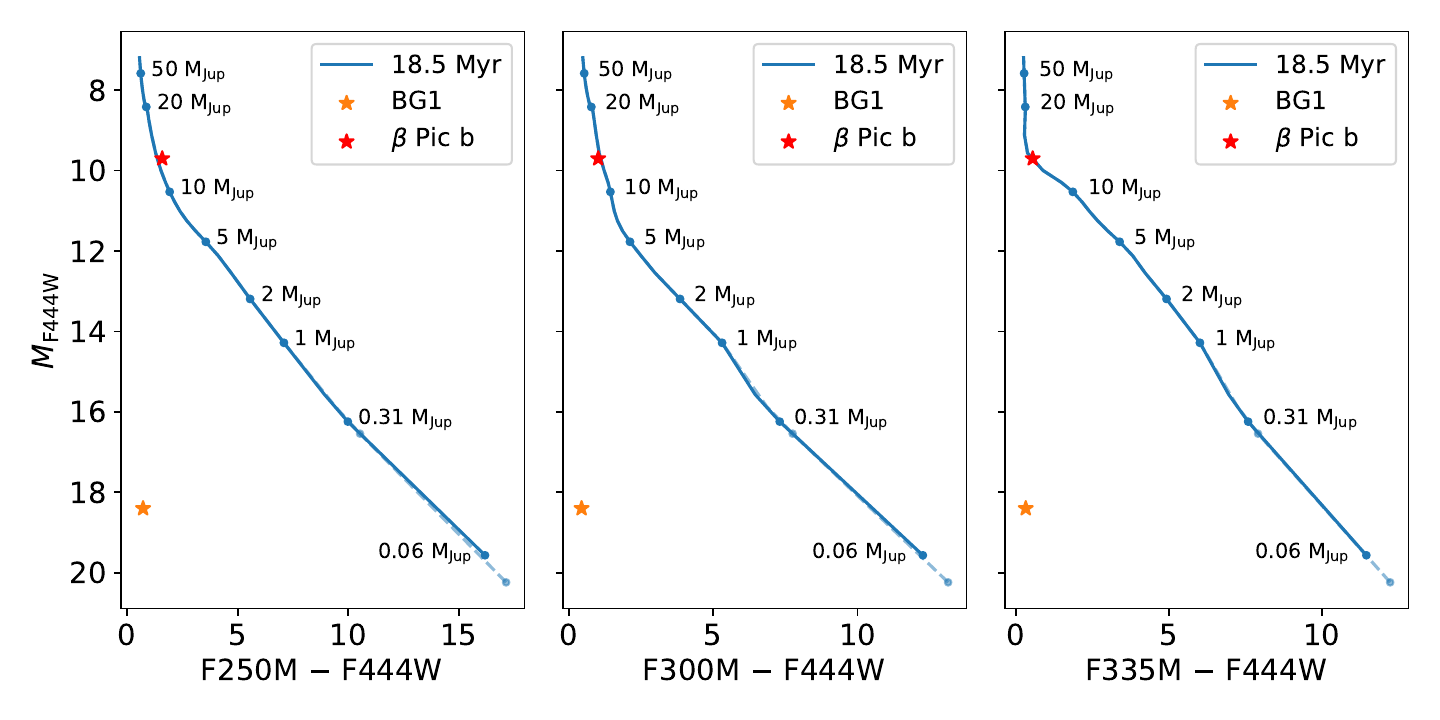}
    \caption{\emph{JWST}/NIRCam color-magnitude diagrams showing an ATMO + BEX hot start evolutionary track for an 18.5~Myr old bound companion in blue. The point-like background source identified in the NIRCam data of \bpic and the giant planet \bpicb are shown by an orange and a red star. The dashed and transparent blue line shows an ATMO + BEX cold start evolutionary track for reference.}
    \label{fig:evotrack_bg1}
\end{figure*}

\begin{figure*}[!t]
    \centering
    \includegraphics[width=\textwidth]{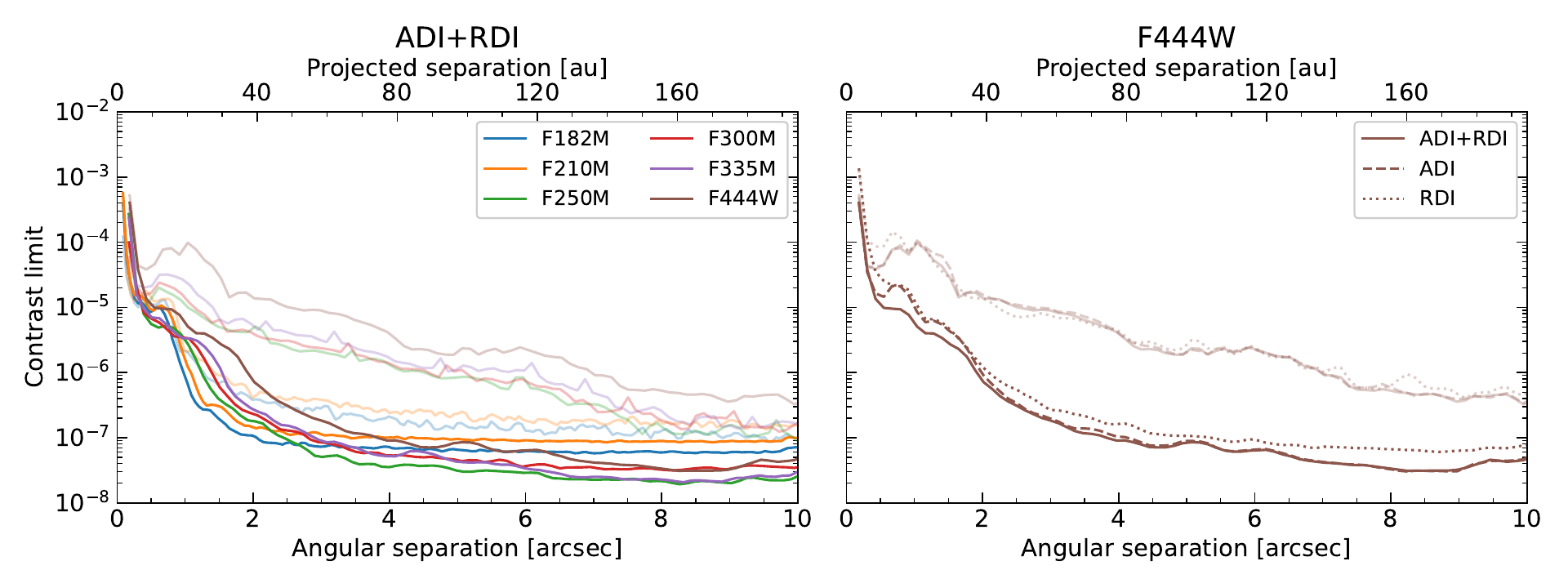}
    \includegraphics[width=\textwidth]{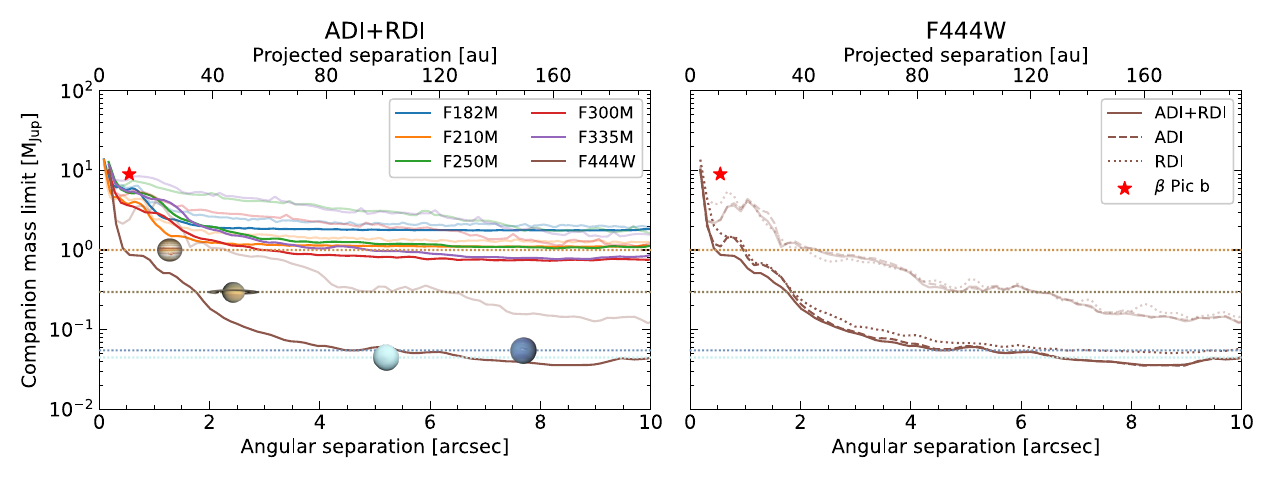}
    \caption{$5\sigma$ contrast limits (top) and companion mass limits (bottom) measured from the PSF-subtracted \emph{JWST}/NIRCam coronagraphy images of the \bpic system. The left panels show the ADI+RDI limits for all six observed filters; the right panels compare the three employed KLIP PSF subtraction techniques (ADI+RDI, ADI, RDI) for the F444W wide-band filter. All curves were computed from high-pass-filtered images to remove the flux of the edge-on debris disk around \bpic as much as possible. The solid lines show the limits away from the disk midplane and the transparent lines show the limits in the disk midplane. The companion mass limits were computed using ATMO + BEX hot start evolutionary models and assuming an age of 18.5~Myr and a distance of 19.44~pc for \bpic. The masses of the gas and ice giants in the Solar System are shown for reference (planet symbol sizes are not to scale, and planet projected separations are five times the true separations from the Sun).}
    \label{fig:detlims}
\end{figure*}

\begin{figure*}[!t]
    \centering
    \includegraphics[width=0.40\textwidth]{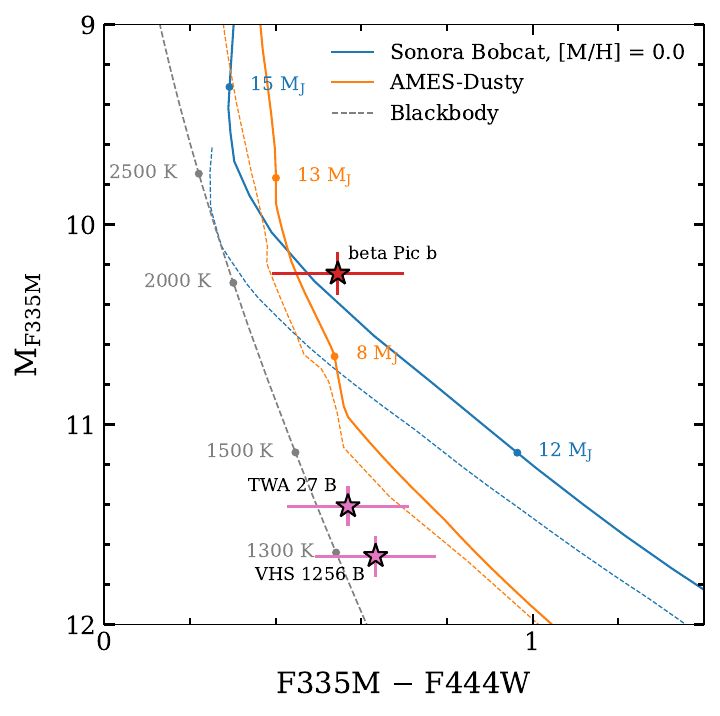}
    \includegraphics[width=0.40\textwidth]{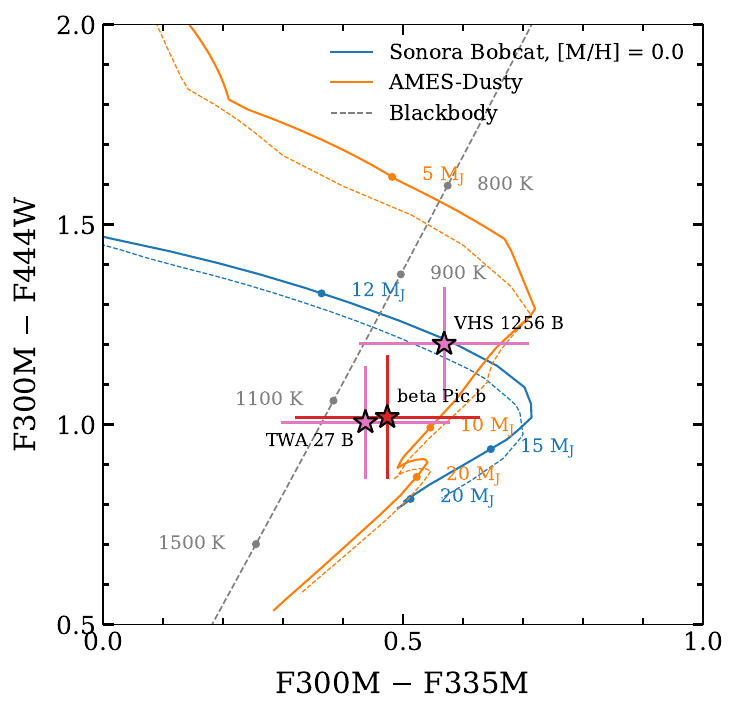}
    \caption{\emph{JWST}/NIRCam color-magnitude diagram (left) and color-color diagram (right) showing Sonora Bobcat and AMES-Dusty isochrones for an 18.5~Myr old companion in blue and orange, respectively. The giant planet \bpicb is shown by a red star and two other substellar companions that were already observed with \emph{JWST} spectroscopy are shown by pink stars. The dashed gray line shows a blackbody and the dashed blue and orange lines show the isochrones for a 100~Myr old companion for reference.}
    \label{fig:atmosphere}
\end{figure*}

To determine the nature of the point-like source (BG1), we extract its photometry by fitting forward-modeled PSFs to the KLIP-subtracted images, similar to how we measured the photometry of \bpicb. Figure~\ref{fig:evotrack_bg1} shows color-magnitude diagrams with the point-like source indicated by an orange star, and an evolutionary track for an 18.5~Myr old bound companion in blue. For reference, \bpicb is also shown by a red star. The evolutionary track was obtained by combining ATMO equilibrium chemistry models from \citet{phillips2020} with BEX hot and cold start models from \citet{linder2019}. This was necessary to cover the entire planet mass range from $\sim0.1$--$13~\Mjup$ to which \emph{JWST} is sensitive. We first interpolated the evolutionary models at an age of 18.5~Myr, and then stitched together the ATMO and BEX models by interpolating linearly between them in log(mass) space. The transition between the two models happens between 0.5 and $1~\Mjup$ where both models are defined. We focused on chemical equilibrium models without clouds to maintain consistency among the different companion mass ranges. We further assumed a distance of 19.44~pc \citep{vanleeuwen2007} for \bpic. The point-like source is far too blue in order to be a young and bound planetary-mass companion. Instead, its colors are more consistent with a background star. We note that we did also inspect archival \emph{HST}/STIS and WFC3 data (program ID 12551, PI: D.~Apai, and 11150, PI: J.~Graham, respectively), but did not identify any background star whose position and proper motion would agree with the point-like source seen in the NIRCam data. Given the position of the source near the edge-on debris disk and the proper motion of \bpic, it seems possible that the source might have been right behind the disk in the archival \emph{HST} images. In any case, its colors indicate it is almost certainly a star.

Due to its unparalleled sensitivity, the NIRCam data also allow us to put deep constraints on the presence of additional companions in the \bpic system. First, contrast curves were computed from the KLIP-subtracted images assuming a noise distribution following a student's t-distribution and correcting for small sample statistics following \citet{mawet2014}. To avoid being impacted by residual flux from the debris disk when computing the contrast curves, we here consider the high-pass filtered data (with a high-pass filter size of 3~pixels). Nevertheless, since some amount of residual noise from the debris disk remains in the high-pass-filtered images, we always show two kinds of detection limits: (1) those in a region far away from the disk and (2) those in a small region around the disk midplane. Then, using the ATMO + BEX hot start evolutionary models introduced in the previous paragraph, the companion mass sensitivity of our observations was derived from the contrast curves. Figure~\ref{fig:detlims} shows the contrast and companion mass limits obtained for the different filters as a function of the angular separation from the host star \bpic. The contrast with respect to \bpic was obtained by comparing the flux measured in the NIRCam images to the flux of \bpic according to a BOSZ A6V-type stellar atmosphere model \citep{bohlin2017} fitted to the archival 1--$5~\um$ photometry of \bpic obtained from VizieR\footnote{\url{https://vizier.cfa.harvard.edu/vizier/sed/}}. The contrast limits reach down to better than $1\mathrm{e}{-7}$ beyond $\sim4$~arcsec ($\sim80$~au). In the background-limited regime, this is about 10 times deeper in the L-band and 30 times deeper in the M-band if compared to the NACO detection limits reported in \citet{stolker2019}. The ADI+RDI and ADI reductions perform fairly similar at close-in and wide separations and there is an intermediate regime ($\sim0.5$--2~arcsec) where ADI+RDI performs better than ADI alone. We note that this regime is where the coronagraphic optics create a strong quasi-static speckle field in the images. The RDI only reduction performs the worst. As discussed in Section~\ref{sec:image_alignment_and_target_acuisition_performance}, this is due to the poor reference star subtraction which resulted from the suboptimal TA performance for the reference star observations of \apic in the LW channel.

Looking at the companion mass limits, the F444W wide-band filter outperforms all other filters by a factor of $\gtrsim10$ at wide separations and reaches a $5\sigma$ detection limit of $\sim0.05~\Mjup$ (approximately a Neptune/Uranus mass) beyond $\sim4$~arcsec ($\sim80$~au) away from the disk midplane. The far deeper sensitivity of the F444W filter has multiple reasons. Firstly, the wide-band filters have a significantly increased throughput compared to the medium-band filters, resulting in higher sensitivity. Secondly, the contrast between a young giant planet and an A-type host star is smaller at longer wavelengths \citep[e.g.,][]{beiler2023,leggett2023}, which benefits the F444W filter the most. We can hence rule out the presence of companions away from the disk midplane with masses above $\sim0.05~\Mjup$ between $\sim4$ and 10~arcsec ($\sim80$--200~au) from \bpic. In the disk midplane, the companion mass limits are worse due to residual photon noise and flux from the disk. We reach a $5\sigma$ sensitivity of $\sim1~\Mjup$ beyond 2~arcsec ($\sim40$~au) from \bpic.



\section{Discussion}
\label{sec:discussion}

\subsection{The atmosphere of \bpicb}
\label{sec:the_atmosphere_of_bpicb}

\begin{table*}[!t]
    \centering
    \begin{tabular}{c c c c c c}
        Model & $T_{\rm{eff}}$ [K] & $\log g$ [dex] & [Fe/H] & C/O & $R$ [$\Rjup$] \\
        \hline
        BT-Settl & 1000--3000 & 3.5--5.5 & -- & -- & 0.5--2.5 \\
        DRIFT-PHOENIX & 1000--3000 & 3.0--5.5 & -0.6--0.3 & -- & 0.5--2.5 \\
        Exo-REM & 1000--2000 & 3.0--5.0 & -0.5--1.0 & 0.1--0.8 & 0.5--2.5 \\
    \end{tabular}
    \caption{Prior ranges adopted for the considered model atmosphere grids. The priors were distributed uniformly within the quoted ranges.}
    \label{tab:priors}
\end{table*}

\begin{table*}[!t]
    \centering
    \begin{tabular}{c c c c c c c}
        Model & $T_{\rm{eff}}$ [K] & $\log g$ [dex] & [Fe/H] & C/O & $R$ [$\Rjup$] & $\log L/L_\odot$ [dex] \\
        \hline
        BT-Settl (all data) & $1640^{+3}_{-4}$ & $3.51^{+0.01}_{-0.01}$ & -- & -- & $1.68^{+0.01}_{-0.01}$ & $-3.71^{+0.00}_{-0.00}$ \\
        BT-Settl (NIRCam only) & $1567^{+96}_{-71}$ & $3.84^{+0.17}_{-0.10}$ & -- & -- & $1.64^{+0.11}_{-0.17}$ & $-3.82^{+0.02}_{-0.03}$ \\
        DRIFT-PHOENIX (all data) & $1723^{+5}_{-4}$ & $4.00^{+0.05}_{-0.04}$ & $0.14^{+0.02}_{-0.01}$ & -- & $1.44^{+0.01}_{-0.01}$ & $-3.76^{+0.00}_{-0.00}$ \\
        DRIFT-PHOENIX (NIRCam only) & $1671^{+60}_{-63}$ & $4.04^{+0.09}_{-0.10}$ & $-0.20^{+0.34}_{-0.30}$ & -- & $1.40^{+0.09}_{-0.08}$ & $-3.84^{+0.03}_{-0.03}$ \\
        Exo-REM (all data) & $1452^{+4}_{-3}$ & $3.76^{+0.03}_{-0.04}$ & $0.80^{+0.05}_{-0.04}$ & $0.37^{+0.01}_{-0.01}$ & $2.03^{+0.02}_{-0.02}$ & $-3.76^{+0.01}_{-0.01}$ \\
        Exo-REM (NIRCam only) & $1450^{+75}_{-62}$ & $3.79^{+0.09}_{-0.10}$ & $0.34^{+0.46}_{-0.52}$ & $0.53^{+0.14}_{-0.22}$ & $1.88^{+0.12}_{-0.12}$ & $-3.82^{+0.04}_{-0.04}$ \\
        \end{tabular}
    \caption{Best fit bulk parameters inferred for the young giant planet \bpicb using different model atmosphere grids. For each model grid, we separately report the results of fits to all available data, and to the NIRCam data only.}
    \label{tab:posteriors}
\end{table*}

\begin{figure*}[!t]
    \centering
    \includegraphics[width=0.70\textwidth]{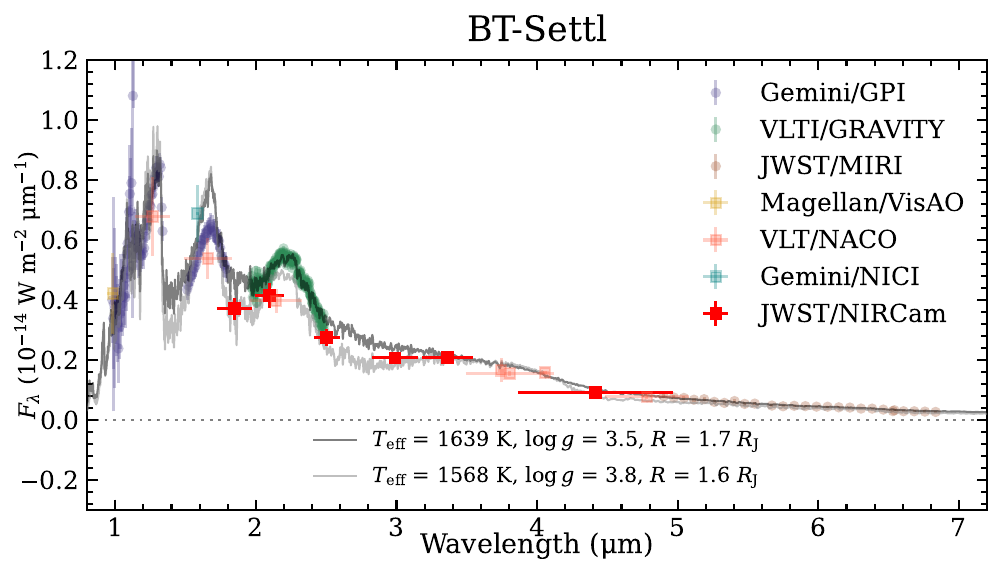}
    \includegraphics[width=0.70\textwidth]{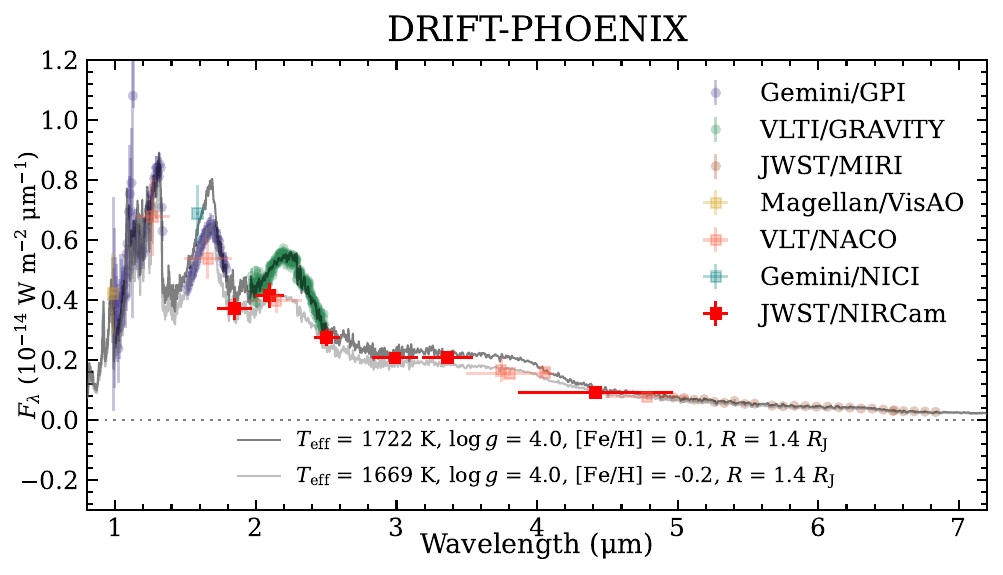}
    \includegraphics[width=0.70\textwidth]{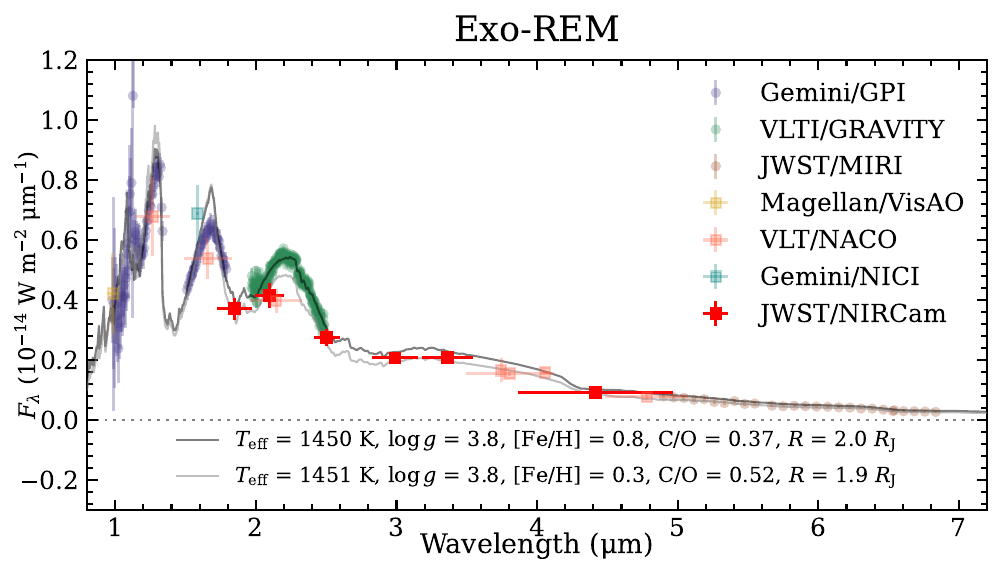}
    \caption{Best fit model atmospheres from the BT-Settl (top), DRIFT-PHOENIX (middle), and Exo-REM (bottom) grids fitted to the \emph{JWST}/NIRCam photometry of the giant planet \bpicb alone (light gray) and together with existing data from the literature (dark gray).}
    \label{fig:spectra}
\end{figure*}

\begin{figure*}[!t]
    \centering
    \includegraphics[width=0.80\textwidth]{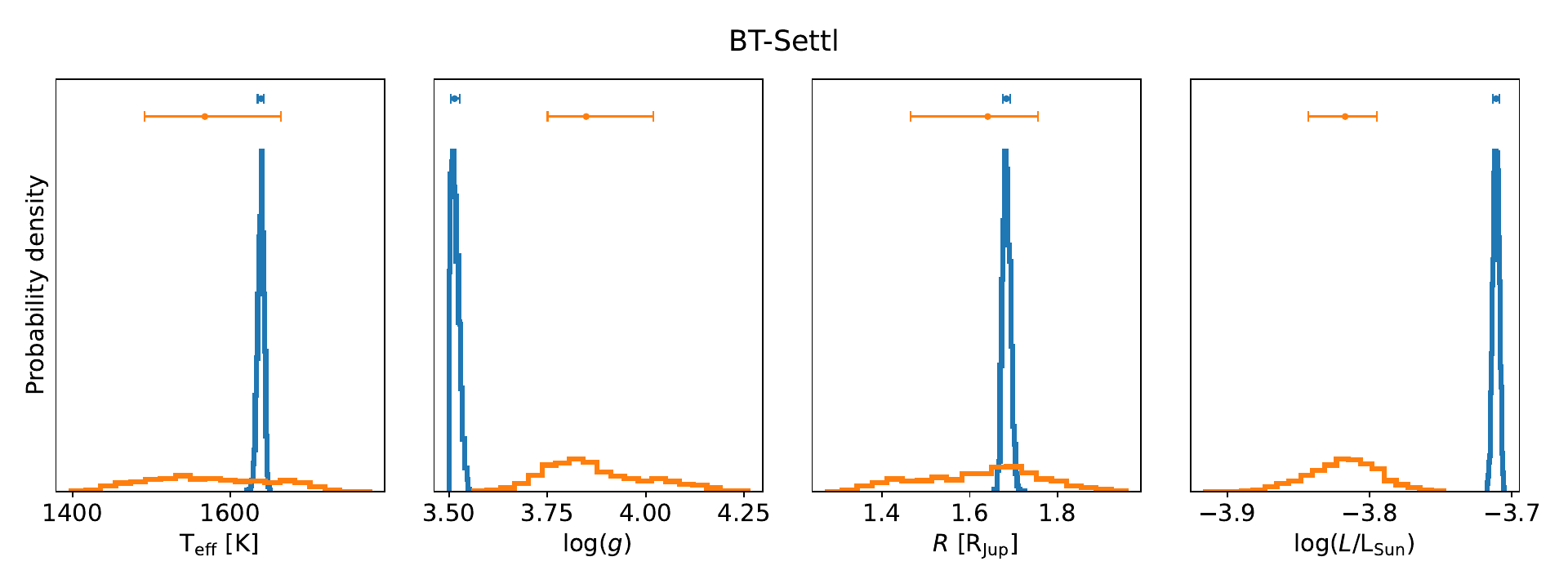}
    \includegraphics[width=0.80\textwidth]{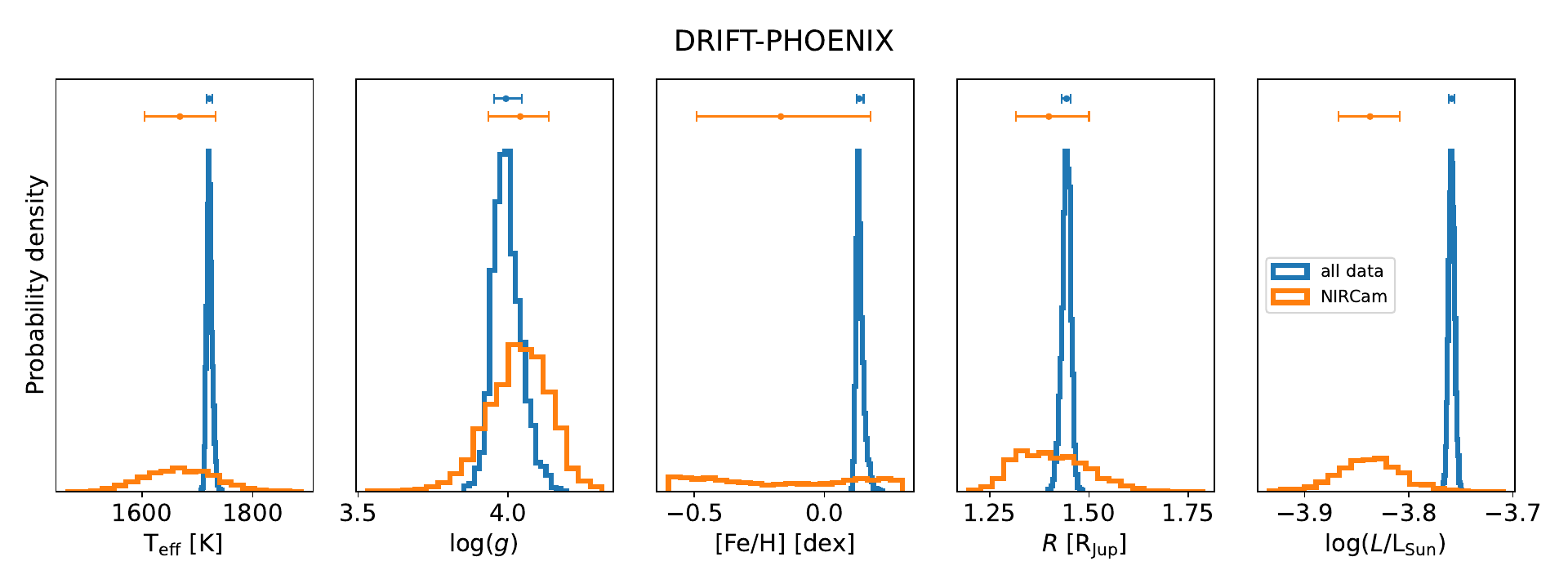}
    \includegraphics[width=0.80\textwidth]{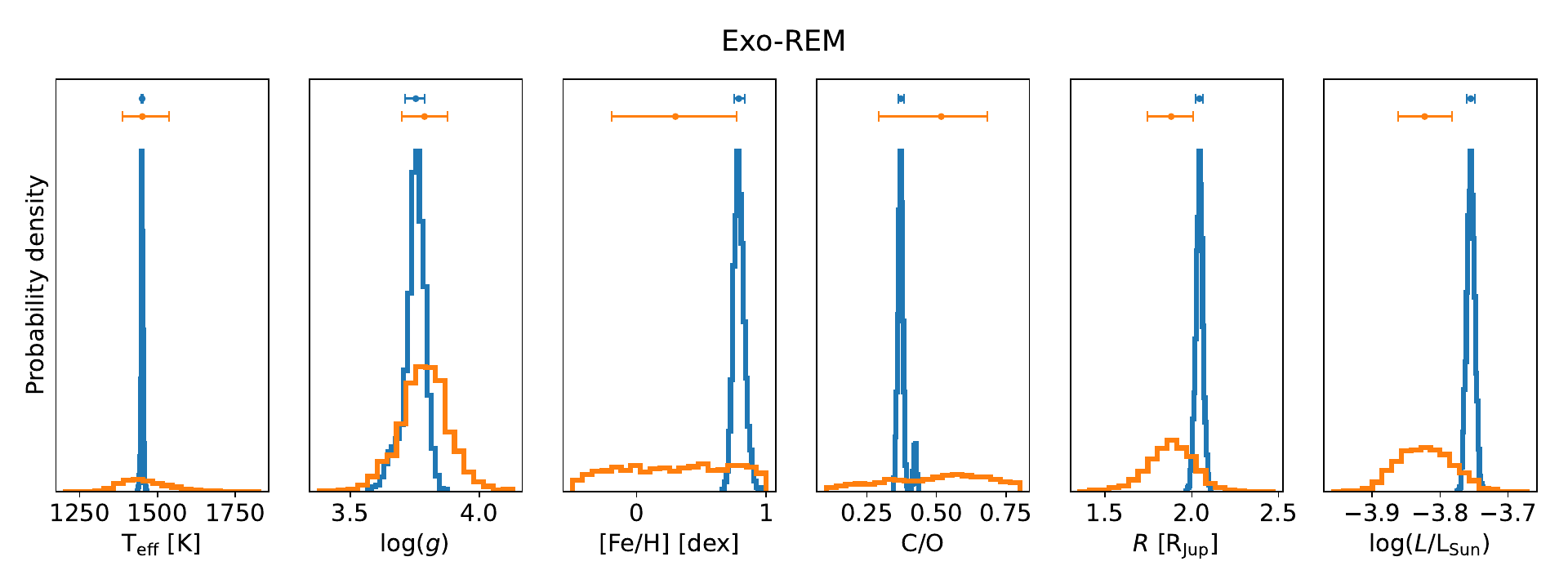}
    \caption{Model atmosphere parameter posterior distributions for the BT-Settl (top), DRIFT-PHOENIX (middle), and Exo-REM (bottom) grids fitted to the \emph{JWST}/NIRCam photometry of the giant planet \bpicb alone (orange) and together with existing data from the literature (blue). The markers with horizontal error bars above the histograms show the 16th, 50th, and 84th percentile of the distribution.}
    \label{fig:posteriors}
\end{figure*}

\begin{figure*}[!t]
    \centering
    \includegraphics[width=\textwidth]{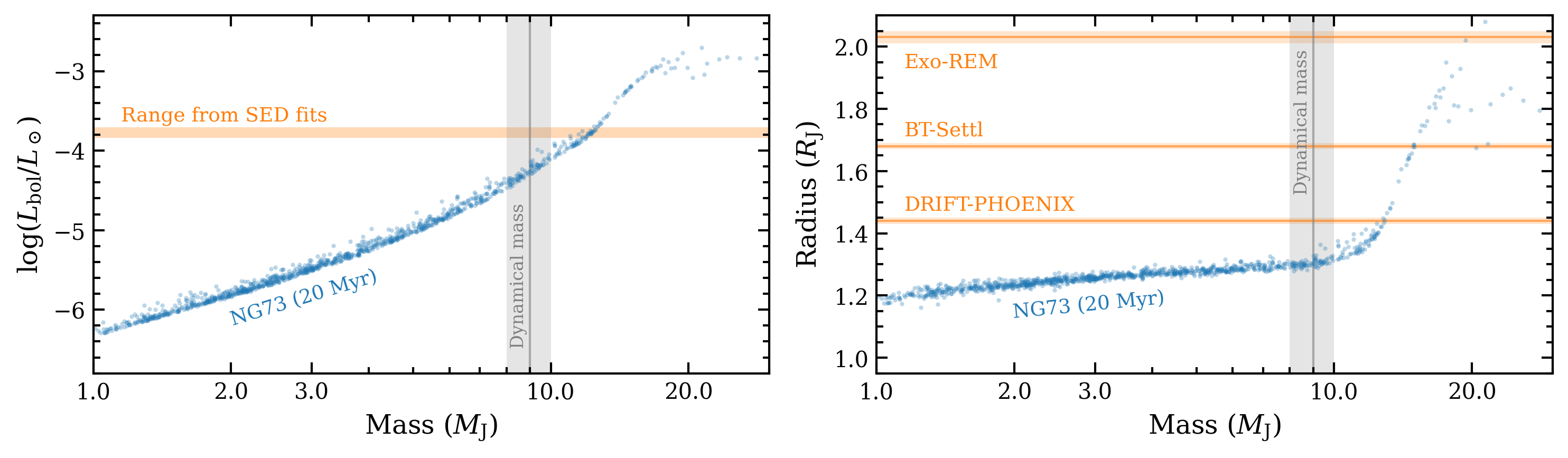}
    \caption{Comparison of \bpicb's inferred luminosity and radius with the New Generation Planetary Population Synthesis \citep[NGPPS;][]{emsenhuber2021}. The left panel shows the bolometric luminosity, with the orange area highlighting the combined constraints from the SED fits (\rev{all six models in Table~\ref{tab:posteriors}}). The right panel shows the radius, with the constraints from the three model atmosphere grids highlighted separately. The samples from NGPPS were selected from \texttt{NG73} (i.e., one planetary embryo per simulated system) at an age of 20~Myr. The gray area shows the dynamical mass of \bpicb, indicating a discrepancy with the bulk constraints from the SED.}
    \label{fig:formation}
\end{figure*}

Figure~\ref{fig:atmosphere} shows the position of \bpicb in a NIRCam color-magnitude and color-color diagram together with AMES-Dusty \citep{allard2001} and Sonora Bobcat \citep{marley2021} isochrones. For reference, two other substellar companions that were already observed with \emph{JWST}/NIRSpec spectroscopy (VHS~1256~B, \citealt{miles2023}, and TWA~27~B, \rev{\citealt{luhman2023,manjavacas2024}}) are also shown. These diagrams can inform the evolution of substellar companions from the correlation of their luminosity and spectral type, and from correlations between the spectral slopes of two different wavelength regimes, respectively. Both diagrams were made with the \texttt{species} toolkit \citep{stolker2020}. Given the dynamical mass of $\sim9\pm2~\Mjup$ of \bpicb \citep{nowak2020,brandtetal2021}, the young giant planet  agrees better with the AMES-Dusty isochrone, which is not surprising given that previous works have already identified \bpicb to have a cloudy atmosphere \citep[e.g.,][]{currie2013,bonnefoy2013,morzinski2015,chilcote2017}. Both isochrones assume equilibrium chemistry which plays a role for the depth of the methane feature discussed below. While Sonora Bobcat describes a cloudless atmosphere with only rainout chemistry included, AMES-Dusty completely neglects the gravitational settling of the grains so that layers of clouds automatically build up through condensation. The dust clouds absorb light at shorter wavelengths and re-emit it in the near-IR, leading to a redder color and thus an increased brightness in the near-IR if compared to a cloudless atmosphere at the same surface gravity (i.e., planet mass). In the color-color diagram, both isochrones predict that objects will become bluer in F300M $-$ F444W with increasing mass, because the effective temperature also increases, moving the blackbody peak to shorter wavelengths. These colors are hence probing mostly the pseudo-continuum of the atmosphere. The combination of the F300M and the F335M filters provides a tracer of the methane absorption feature at $\sim3.3~\um$. The F300M $-$ F335M color first becomes redder at small masses, because the depth of the methane absorption feature in the F335M filter decreases. Then, at a mass of about $\sim7~\Mjup$ for AMES-Dusty and $\sim14~\Mjup$ for Sonora Bobcat, the methane feature completely disappears and objects become bluer with increasing mass as expected from a pure blackbody. It can be seen that the giant planet \bpicb is already in the regime where the methane feature has disappeared. We note, however, that the evolutionary models used here do not include disequilibrium chemistry so that methane absorption already appears at higher effective temperatures compared to what has been observed \citep{yamamura2010,zahnle2014,moses2016,stolker2020}. For a low-gravity L-type object such as \bpicb, methane absorption is not expected.

To derive a set of physical parameters of \bpicb based on our \emph{JWST}/NIRCam photometry, we fitted model atmospheres to its observed spectrophotometry using the \texttt{species} toolkit \citep{stolker2020}. Besides the new \emph{JWST}/NIRCam photometry, we included the Gemini/GPI spectra from \citet{chilcote2017}, the VLTI/GRAVITY spectrum from \citet{gravity2020}, the recent \emph{JWST}/MIRI spectrum from \citet{worthen2024}, the Magellan/VisAO and Gemini/NICI photometry from \citet{males2014}, and the VLT/NACO photometry from \citet{bonnefoy2011}, \citet{currie2013}, \citet{stolker2019}, and \citet{stolker2020}. We explored three different model atmosphere grids: the DRIFT-PHOENIX \citep{helling2008}, the Exo-REM \citep{baudino2015,charnay2018}, and the BT-Settl \citep{allard2012} grid. All of them assume radiative-convective equilibrium to calculate the temperature structure of the atmosphere and include photospheric absorption by dust clouds. Cloudy model atmospheres have been found to provide good fits to the observed spectrophotometry of \bpicb by many previous works \citep[e.g.,][]{currie2013,bonnefoy2013,stolker2020,gravity2020}. A cloudy atmosphere is also expected for \bpicb based on its effective temperature and the condensation temperatures of common refractory species in its atmosphere \citep[e.g.,][]{marley2015}. To be able to fit the different model grids to the data, we first bin them to the spectral resolution of the respective instrument and calculate synthetic photometry for all considered filters. Then, we interpolate the binned spectra and synthetic photometry linearly between the different grid points and infer the posterior distribution of the grid parameters using nested sampling with the \texttt{PyMultiNest}\footnote{\url{https://github.com/JohannesBuchner/PyMultiNest}} package \citep{buchner2014}. Uniform distributions were chosen for the priors in the parameter ranges shown in Table~\ref{tab:priors}.

To investigate the characterization power of NIRCam by itself, we perform two types of fits: (1) using only the new NIRCam photometry and (2) including the existing literature data in addition to the new NIRCam photometry. Given that \bpicb is such a well-studied object, these tests are ideally suited to check which atmospheric parameters our chosen set of NIRCam filters can constrain by themselves and which not. This knowledge will be especially interesting for future NIRCam observations of previously unknown exoplanetary companions. The best fit parameter values for the three considered model atmosphere grids and the two types of fits are shown in Table~\ref{tab:posteriors}. \rev{We note that fits using only the existing literature data without the new NIRCam photometry do not significantly differ from the ones including the new NIRCam photometry and are hence not shown or further discussed here. There are multiple reasons for this. Firstly, the GPI and GRAVITY spectra have so many data points that the weight of the NIRCam photometry is very small. Secondly, there is already NACO photometry in the L- and M-bands and the error bars of the NIRCam photometry are too large to further constrain the best fit models in this wavelength regime.}

Figure~\ref{fig:spectra} shows the best fit model atmospheres for \bpicb for all six considered cases. The fits that consider only the new NIRCam photometry favor an object which emits less flux over the shown spectral range and has a smaller bolometric luminosity. Notably, the NACO photometry and the GPI H-band spectrum appear to fit better to the model atmosphere derived from the NIRCam photometry alone. This means that the NACO and GPI H-band data seem to be more consistent with the new NIRCam photometry than with the existing GRAVITY spectrum, which pulls the entire near-IR spectrum of \bpicb up to higher fluxes in the fits that include the literature data. We note that these fits use an equal weight for each data point, so that the GRAVITY spectrum, although correlations between spectral channels are taken into account, dominates the fit. Investigating the origin of the systematic differences between the GRAVITY spectrum and the other instruments is beyond the scope of this paper, but we note in Section~\ref{sec:jwst_photometry} that variability in the scattered light from the debris disk could play a role.


Comparing, for each model, the fits with the NIRCam data alone to the ones including existing literature data, Figure~\ref{fig:posteriors} shows that the retrieved posteriors for the effective temperature, the surface gravity, and the radius are mostly consistent with one another within $1\sigma$. As observed frequently in the literature, there are however systematic differences between the different model grids. In general, the retrieved effective temperature ranges from $\sim1450$--1750~K, which is consistent with previous studies \citep[e.g.,][]{bonnefoy2014,gravity2020}. However, it becomes evident that at the current photometric precision our specific set of NIRCam filters is not suitable to constrain the planet's metallicity or C/O abundance ratio. Figure~\ref{fig:posteriors} illustrates that the posterior distributions of these quantities remain unconstrained regardless of the chosen model atmosphere grid. When including literature data in the fits, the metallicity and C/O abundance ratio posterior distributions converge sharply. With the Exo-REM grid, we retrieve a C/O abundance ratio of $0.37^{+0.01}_{-0.01}$ which is consistent with the value retrieved in \citet{gravity2020} with the GRAVITY K-band spectrum alone. The retrieved metallicity is strongly model-dependent. \rev{We note that we can obtain much tighter constraints on the metallicity and C/O abundance ratio from our NIRCam observations alone when artificially increasing the photometric precision to $\sim1\%$. When doing so, the uncertainty on the retrieved metallicity drops to $\sim0.05$~dex and the uncertainty on the retrieved C/O ratio drops to $\sim0.1$. We suspect that the six photometric points from NIRCam alone can constrain these quantities quite tightly because they cover such a broad wavelength range ($\sim1.7$--$5~\um$). However, similar as above, there remain systematic differences between the different model grids so that the retrieved parameter values need to be treated with caution.}

For all three model grids, though, we find that the luminosity retrieved from the NIRCam photometry alone is significantly smaller than the one retrieved from the fit including literature data. This is not completely surprising given that the new NIRCam photometry falls consistently below the best fit DRIFT-PHOENIX model atmosphere fitted to literature data shown in Figure~\ref{fig:photometry}. Our specific set of NIRCam filters covers both the water bands at $\sim2$ and $\sim2.5~\um$ and the CO/CO${}_2$ bands in the $\sim4$--$5~\um$ regime, which are the main carbon- and oxygen bearing species in the atmosphere of \bpicb. However, we are lacking medium band observations between 4--$5~\micron$ which could provide a better handle on the relative strength of the CO and CO${}_2$ features.

\subsection{The formation of \bpicb}
\label{sec:the_formation_of_bpicb}

The formation history of \bpicb has been investigated in several studies, and most of them concluded that either its luminosity-age relationship \citep[e.g.,][]{bonnefoy2014,marleau2014,nowak2020} or its atmospheric composition \citep[e.g.,][]{gravity2020} are most consistent with a moderate formation entropy ($\gtrsim9.75~k_{\rm{B}}/\rm{baryon}$) and suggesting a formation by warm-start core-accretion \citep{pollack1996,spiegel2012,mordasini2013} followed by strong planetesimal enrichment in the inner regions of the \bpic system \citep[i.e., within the CO${}_2$ iceline, e.g.,][]{oeberg2011,eistrup2018}. Based on the new \emph{JWST}/NIRCam photometry which improves the constraints on \bpicb's thermal emission especially in the $\sim3$--$5~\um$ regime, we obtain slightly updated constraints on the planet's bolometric luminosity and radius. We note that while the $\sim5$--$7~\micron$ MIRI MRS spectrum of \bpicb is included in the fits, it does not contribute significantly to the estimated bolometric luminosity of the planet because we let the spectrum scale up or down freely in these fits, following the same procedure as in \citet{worthen2024}. \rev{Moreover, as in \citet{stolker2020}, we keep the GRAVITY and GPI Y-J-band spectra fixed while also fitting a scale factor for the GPI H-band spectrum which does otherwise not agree with the GRAVITY spectrum. A similar discrepancy between the GPI H-band spectrum and the GRAVITY and SPHERE spectra of the brown dwarf companion HD~206893~B has been reported by \citet{kammerer2021}.}

We compare these updated values to evolutionary model predictions from the New Generation Planetary Population Synthesis (NGPPS) simulation \citep{emsenhuber2021} in Figure~\ref{fig:formation}. The NGPPS simulation uses a global end-to-end planetary formation and evolution model based on the core-accretion process to predict all common observational quantities of exoplanets and exoplanetary systems. It can be seen that the dynamical mass from \citet{nowak2020} and  \citet{brandtetal2021} is in tension with the luminosity measured from the SED fits which suggests that \bpicb has a mass closer to the deuterium burning limit. This was already noted by \citet{brandtetal2021}. The same conclusion can be drawn from the inferred planet radius, which (regardless of the used model atmosphere grid) is also in tension with the one inferred from the NGPPS and the dynamical mass of \bpicb. Investigating the reason for this discrepancy is beyond the scope of this paper. However, we note that \citet{lacour2021} find a more consistent dynamical mass of $11.9\pm3~\Mjup$ for \bpicb when only considering the radial velocities and the astrometry, ignoring the Gaia-Hipparcos proper motion anomaly.


Our inferred bolometric luminosity is consistent with the one reported in \citet{morzinski2015} and \citet{chilcote2017} to within $\pm0.05$~dex, however, the comparison of dynamical mass and NGPPS predicts a bolometric luminosity that is $\sim0.5$~dex lower. We further note that the radius inferred by the Exo-REM grid is very large, almost unphysical if compared to the NGGPS population. However, it is roughly consistent with the radius of $1.9~\Rjup$ inferred by \citet{gravity2020} also using the Exo-REM grid. With the free retrieval code \texttt{petitRADTRANS}, the same authors obtain a radius of $1.36\pm0.02~\Rjup$ which is much closer to the one that we infer using the DRIFT-PHOENIX model atmosphere grid. The large discrepancies between the inferred radii can originate from the different cloud prescriptions in the different model atmosphere grids and retrieval codes. Clouds can have a similar effect on a planet's SED as a change in effective temperature and radius and are therefore often correlated with these quantities. In the literature, the DRIFT-PHOENIX grid and the cloud prescription therein has been found to provide good fits to young and dusty low-gravity objects \citep{patience2012,bonnefoy2014a,lachapelle2015} such as \bpicb \citep{stolker2020}, HD~95086~b \citep{derosa2016}, and HD~206893~B \citep{kammerer2021}.

\section{Summary and conclusions}
\label{sec:summary_and_conclusions}

\begin{itemize}
    \item We observed the \bpic exoplanetary system with \emph{JWST}/NIRCam coronagraphy. We detected the directly-imaged giant planet \bpicb in all six observed NIRCam filters.
    \item We measure new photometry for \bpicb which mostly agrees with existing data from the ground. We observe a $\sim10$--20\% discrepancy with respect to the GRAVITY spectrum in the K-band. We note that a similar discrepancy can be seen between the NACO and the GRAVITY data, and that our NIRCam photometry is consistent with NACO within the uncertainties. As it was found in previous studies, our new NIRCam photometry supports that \bpicb has an atmosphere composed of a thick layer of clouds.
    \item We measure new astrometry for \bpicb which is still affected by significant systematic errors. Given the unprecedented accuracy of the existing VLTI/GRAVITY astrometry, our new NIRCam astrometry does not provide updated constraints on the orbital parameters or the dynamical mass of \bpicb. Making the \emph{JWST} astrometry competitive with measurements from single-dish ground-based telescopes requires further work. While the implementation of an updated distortion correction is already being worked on by the instrument and \texttt{spaceKLIP} teams, the largest systematic error is introduced by the unknown position of the host star behind the coronagraphic mask. Observers who care about accurate companion astrometry are hence encouraged to take astrometric confirmation full frame images and use nearby field stars to interpolate the position of the occulted host star behind the coronagraphic mask.
    \item Thanks to the unparalleled infrared sensitivity of \emph{JWST} especially in the NIRCam F444W filter, we are able to rule out additional companions in the disk midplane above $\sim1~\Mjup$ beyond 40~au from the star. Away from the disk midplane, the companion mass limits reach down to $\sim0.05~\Mjup$ beyond 80~au from the star. In particular, we can rule out companions more massive than $\sim1~\Mjup$ in the disk midplane and $\sim0.1~\Mjup$ away from the disk midplane outward of the location of the southwestern clump in the debris disk \citep[e.g.,][]{han2023,skaf2023,rebollido2024}. We further identify five additional localized sources in the NIRCam LW data, but all of them are found to be background stars or galaxies based on their colors or spatial extent.
    \item By fitting model atmosphere grids to the new NIRCam photometry alone and combined with existing data from the literature, we determine the physical and atmospheric parameters of \bpicb. We found that our NIRCam observations are not able to constrain the metallicity and C/O abundance ratio of the planet. However, the effective temperature, surface gravity, and radius inferred from the NIRCam only fits are typically consistent with the ones inferred from existing literature data within the uncertainties. The bolometric luminosity inferred from the NIRCam only fits is slightly smaller than the one inferred from existing literature data.
    \item The presence of the extended debris disk around \bpic makes the extraction of accurate planet photometry for \bpicb challenging. We used fake companion injection and recovery tests to address this issue, but these assume that the debris disk and residual speckles are symmetric with respect to the star. Ultimately, we used MCRDI techniques to simultaneously fit for the disk and the planet, but also the accuracy of this approach depends on how accurately the MCRDI disk model is able to resemble the data. These complications illustrate that measuring accurate planet photometry in the presence of an exozodiacal dust disk in the era of the Habitable Worlds Observatory is a potential problem that deserves further attention \citep[e.g.,][]{kammerer2022b,currie2023}.
    \item The known inner planet \bpicc remains undetected behind the occulting spot of the NIRCam coronagraphs. The direct detection of \bpicc with \emph{JWST} would require the use of high-resolution imaging techniques such as NIRISS Aperture Masking Interferometry \citep[AMI,][]{sivaramakrishnan2023}, although even for AMI this planet is challenging as its contrast is expected to be close to the reported limit of $\sim1\mathrm{e}{-4}$ in the L- and M-band. We note that NIRISS Kernel Phase Interferometry \citep[KPI,][]{kammerer2023} is not feasible for this target because \bpic would highly saturate the detector.
\end{itemize}

\begin{acknowledgments}
The authors are grateful for the many efforts of the \emph{JWST} observatory operations teams, in particular for the flight operations team members who responded to the spacecraft anomaly and payload safing event which occurred during the first attempt at these NIRCam observations, diagnosed the problem, and returned \emph{JWST} to science operations in the midst of the holiday season and shortly before the one year anniversary of its launch. 
We thank Martha Boyer for discussions on NIRCam photometric calibration and her critical analyses to improve that calibration, and thank Erik Mamajek for insightful discussions on the discrepant proper motion of \apic. 
This paper reports work carried out in the context of the \emph{JWST} Telescope Scientist Team (\url{https://www.stsci.edu/~marel/jwsttelsciteam.html}) (PI M.~Mountain). Funding is provided to the team by NASA through grant 80NSSC20K0586.
Based on observations with the NASA/ESA/CSA \emph{JWST}, obtained at the Space Telescope Science Institute, which is operated by AURA, Inc., under NASA contract NAS 5-03127.
The data presented in this article were obtained from the Mikulski Archive for Space Telescopes (MAST) at the Space Telescope Science Institute. The specific observations analyzed can be accessed via \dataset[DOI 10.17909/8yq1-qv46]{https://www.doi.org/10.17909/8yq1-qv46}. These observations are associated with \emph{JWST} program 1411 (PI C.~Stark).
I.~R. is supported by grant FJC2021-047860-I and PID2021-127289NB-I00 financed by MCIN/AEI/10.13039/501100011033 and the European Union NextGenerationEU/PRTR. \rev{The manuscript was substantially improved following helpful comments from an anonymous referee.}
\end{acknowledgments}

\vspace{5mm}
\facilities{JWST}
\software{Matplotlib \citep{hunter2007}, NumPy \citep{harris2020}, SciPy \citep{virtanen2020}, Astropy \citep{astropy2013,astropy2018,astropy2022}, jwst \citep{bushouse2023}, spaceKLIP \citep{kammerer2022,carter2023}, WebbPSF \citep{perrin2014}, emcee \citep{foreman-mackey2013}, orbitize! \citep{blunt2020}}

\appendix

\section{Fake companion injection \& recovery tests}
\label{sec:fake_companion_injection_recovery_tests}

Both the directly-imaged giant planet \bpicb as well as the debris disk around \bpic are well detected in all of our \emph{JWST}/NIRCam coronagraphy images (see, e.g., Figure~\ref{fig:klip_subtraction}). To better understand the impact of the debris disk on the KLIP photometry of the planet, we conduct fake companion injection and recovery tests. The idea of such tests is to inject a forward-modeled PSF at a known flux and separation (the fake companion) into the NIRCam data and try to recover it using the same routines that have also been used to extract the astrometry and photometry of \bpicb. Since we forward model the companion PSFs, they already take into account KLIP algorithm throughput losses and companion self-subtraction due to ADI. However, residual flux from the debris disk and remaining stellar speckles may still affect the measured companion properties. Since the debris disk around \bpic is highly symmetric in the NIRCam bands \citep[see][]{rebollido2024}, the ideal position to inject the fake companion is at the same angular separation as \bpicb, but on the opposite side of the star. This allows us to study the impact of the disk on the measured fake companion properties in an environment that is very similar to that of the true companion \bpicb. The flux at which we inject the fake companion is always the flux that was measured for \bpicb from the MCRDI reductions.

\begin{figure*}[!t]
    \centering
    \includegraphics[width=0.32\textwidth]{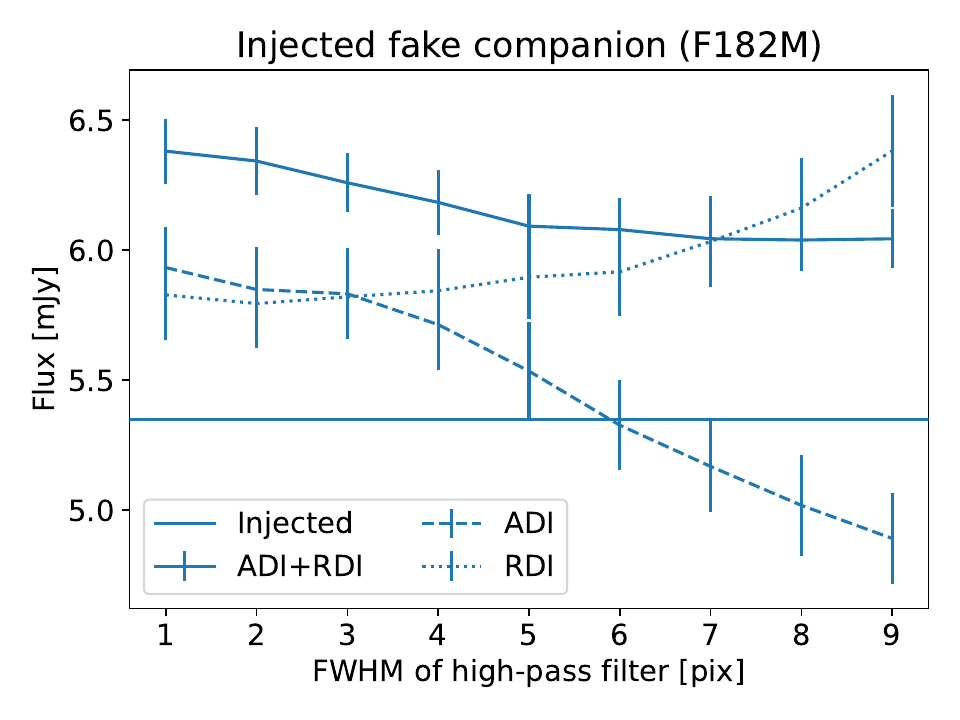}
    \includegraphics[width=0.32\textwidth]{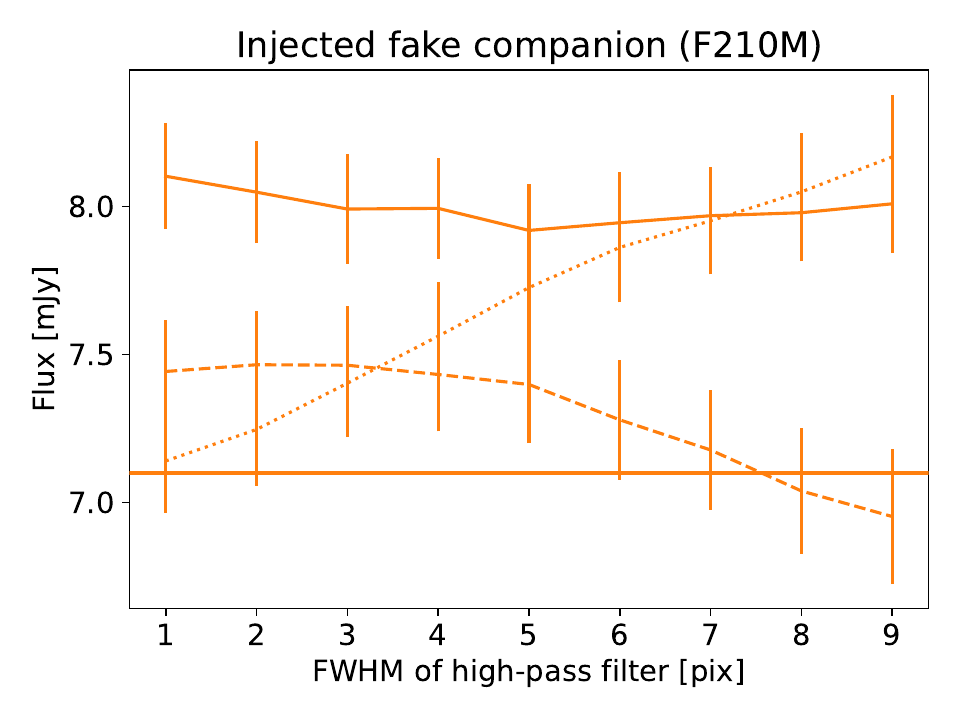}
    \includegraphics[width=0.32\textwidth]{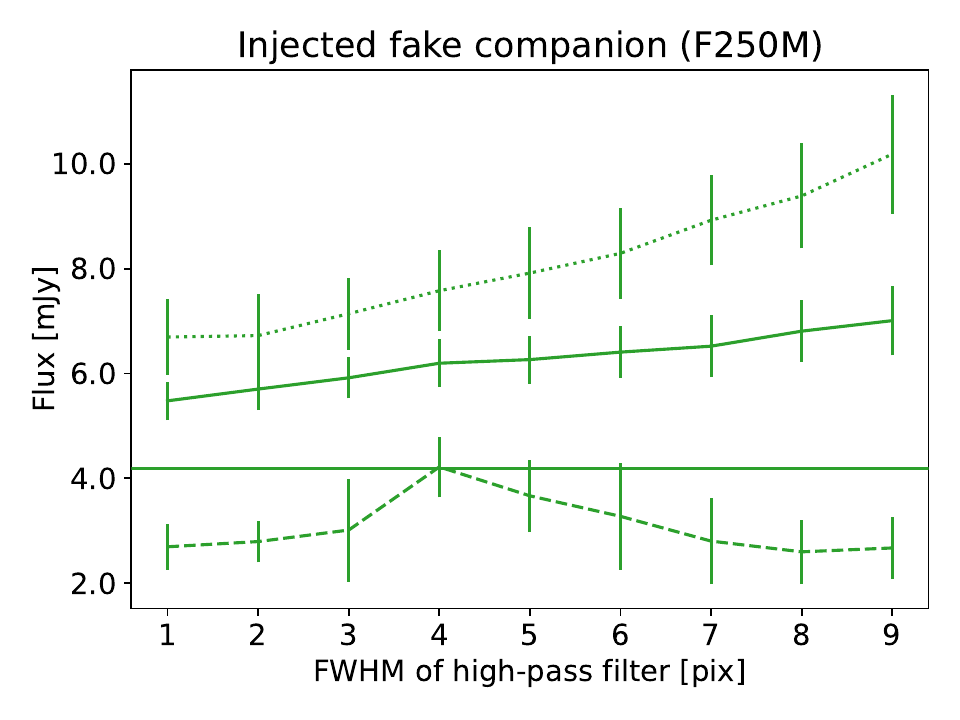}
    \includegraphics[width=0.32\textwidth]{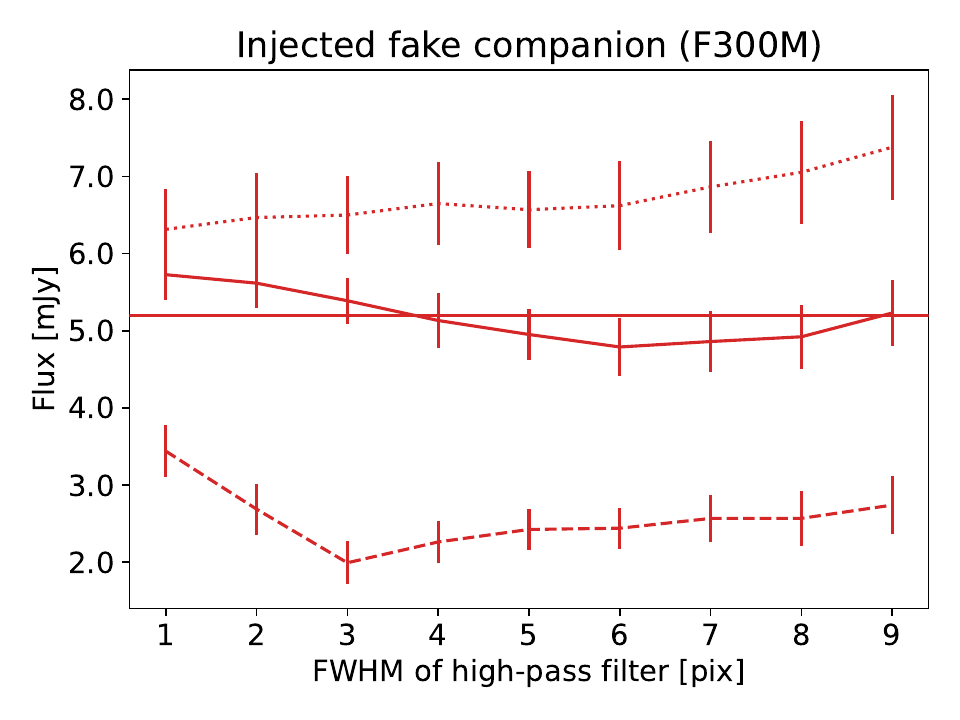}
    \includegraphics[width=0.32\textwidth]{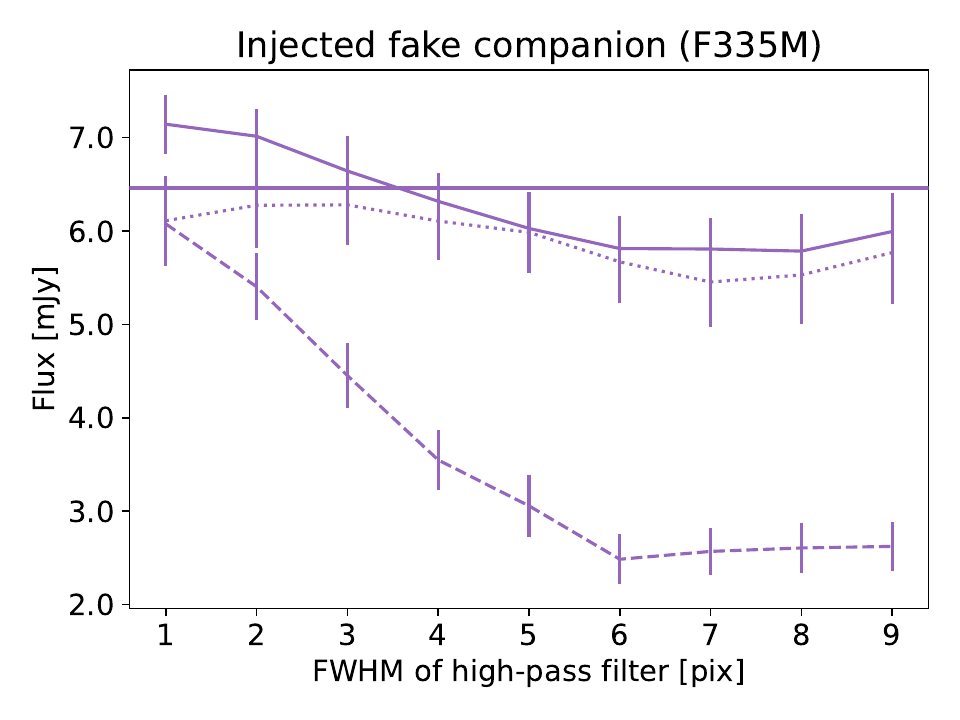}
    \includegraphics[width=0.32\textwidth]{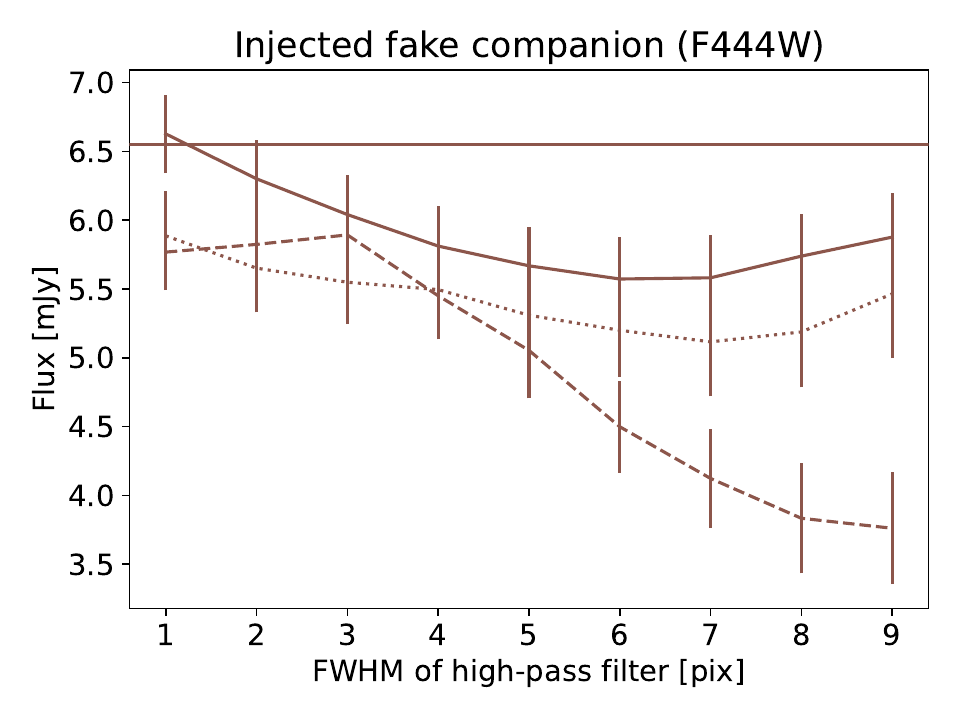}
    \caption{Recovered companion fluxes as a function of the high-pass filter size from fake companion injection and recovery tests using ADI+RDI (solid lines), ADI (dashed lines), and RDI (dotted lines) techniques. The injected fluxes (ground truth) are shown by a horizontal solid line. The fake companion was injected at the same angular separation as \bpicb, but on the opposite side of the star.}
    \label{fig:injection_recovery}
\end{figure*}

Figure~\ref{fig:injection_recovery} shows the results from these fake companion injection and recovery tests using KLIP ADI+RDI, ADI, and RDI techniques. With RDI alone, the recovered companion flux without high-pass filtering is typically higher than the injected flux, because flux from the debris disk adds on top of the companion flux and positively biases the companion flux. With ADI alone, it is the other way around, because disk self-subtraction adds on top of the planet flux and negatively biases the companion flux. If high-pass filtering is included, the recovered companion flux often depends strongly on the high-pass filter size. RDI alone yields either too high or too low flux. ADI alone yields reasonable flux in the SW channel, but too low flux in the LW channel. ADI+RDI yields too high flux in the SW channel, but reasonable flux in the LW channel. Altogether, we adopt an ideal high-pass filter size of 7~pixels for ADI in the SW channel and of 3~pixels for ADI+RDI in the LW channel. While we could in principle choose a different high-pass filter size for each NIRCam filter, this would unnecessarily complicate the analysis given that the recovered fluxes for our adopted parameters are consistent with the recovered fluxes for the ideal high-pass filter sizes within the uncertainties.

\section{Literature photometry of \bpicb}
\label{sec:literature_photometry_of_bpicb}

Table~\ref{tab:literature_photometry} summarizes the literature photometry and spectra of \bpicb used for model atmosphere fitting in this work. We also quote the epochs at which the data has been observed. This might be relevant if variability in the debris disk affects the measured planet fluxes.

\begin{table*}[!t]
    \centering
    \begin{tabular}{c c c c}
        Filter & Flux [mag] & Epoch [YYYY-MM-DD] & Reference \\
        \hline
        Magellan/VisAO.Ys & $15.53\pm0.34$ & 2012-12-04 & \citet{males2014} \\
        Gemini/NICI.ED286 & $13.18\pm0.15$ & 2011-10-20 & \citet{males2014} \\
        Paranal/NACO.J & $14.11\pm0.21$ & 2011-12-16 & \citet{currie2013} \\
        Paranal/NACO.H & $13.32\pm0.14$ & 2012-01-11 & \citet{currie2013} \\
        Paranal/NACO.Ks & $12.64\pm0.11$ & 2010-03-20/2010-04-10 & \citet{bonnefoy2011} \\
        Paranal/NACO.NB374 & $11.25\pm0.26$ & 2012-12-18 & \citet{stolker2020} \\
        Paranal/NACO.Lp & $11.30\pm0.06$ & 2013-02-01 & \citet{stolker2019} \\
        Paranal/NACO.NB405 & $10.98\pm0.04$ & 2012-12-16 & \citet{stolker2020} \\
        Paranal/NACO.Mp & $11.10\pm0.12$ & 2012-11-26 & \citet{stolker2019} \\
        Gemini/GPI.Y & -- & 2015-12-05 & \citet{chilcote2017} \\
        Gemini/GPI.J & -- & 2013-12-10 & \citet{chilcote2017} \\
        Gemini/GPI.H & -- & 2013-11-18/2013-12-10\&11/2014-11-08 & \citet{chilcote2017} \\
        Paranal/GRAVITY & -- & 2018-09-22 & \citet{gravity2020} \\
        JWST/MIRI.MRS & -- & 2023-01-11 & \citet{worthen2024} \\
    \end{tabular}
    \caption{Literature photometry and spectra of the giant planet \bpicb used in this work for model atmosphere fitting.}
    \label{tab:literature_photometry}
\end{table*}

\bibliography{references}{}
\bibliographystyle{aasjournal}

\end{document}